\documentclass[12pt, letterpaper]{article}
\usepackage{amsmath, amssymb,color, natbib, graphicx, caption, subfig, titlesec, setspace, authblk, epstopdf}
\usepackage[letterpaper, margin=1in]{geometry}
\pdfoutput=1

\titleformat*{\section}{\large\bfseries}
\title{Functional regression approximate Bayesian computation for Gaussian process \\ density estimation}
\date{}
\author{G. S. Rodrigues\footnote{School of Mathematics and Statistics, University of New South Wales, Sydney 2052 Australia.}\:\,\footnote{CAPES Foundation, Ministry of Education of Brazil, Bras\'{i}lia - DF 70040-020, Brazil}\:\,\footnote{Communicating Author: {\tt g.rodrigues@unsw.edu.au}}\, ,
David J. Nott\footnote{Department of Statistics and Applied Probability, National University of Singapore, Singapore 117546.}\:\:  and S. A. Sisson$^{*}$}

\begin{document}
\maketitle

\begin{abstract} 
\noindent We propose a novel Bayesian nonparametric method for hierarchical modelling on a set of related density functions, where grouped data in the form of samples from each density function are available. Borrowing strength across the groups is a major challenge in this context. To address this problem, we introduce a hierarchically structured prior, defined over a set of univariate density functions, using convenient transformations of Gaussian processes. Inference is performed through  approximate Bayesian computation (ABC), via a novel functional regression adjustment.  The performance of the proposed method is illustrated via a simulation study and an analysis of rural high school exam performance in Brazil.
\\

\end{abstract}
  
\section{Introduction}
\label{sec:introduction}

We introduce a new statistical procedure for hierarchical modelling on a set of related densities $f_i(x)$ for $i=1, \ldots, g$, based on random samples $\{X_{ij}\}$ such that $X_{ij} \sim f_i(x)$ for $j=1,\dots,n_i$.  The benefits of hierarchical modelling are well known in the parametric context, where the same model is fitted to different but related datasets or groups, and where model parameters are allowed to vary across groups \citep{gelman+csr04}.  A hierarchical prior may be defined on the group varying parameters, and the data used to determine how much pooling of information to perform across groups.  The problem considered here is the nonparametric equivalent of this. Here, the nonparametric density functions $f_i(x)$ are thought to be related and we aim to share information hierarchically to improve estimation of each density, especially for those densities $f_i(x)$ for which the corresponding sample size $n_i$ is small.  

Bayesian nonparametric methods have made enormous advances over the last two decades. The Dirichlet process (DP) \citep{Ferguson1973} has played a central role in this development.  Methods based on Dirichlet process mixture (DPM) models, where a mixing distribution is given a Dirichlet process prior, are a standard approach to flexible Bayesian density estimation  \citep{Lo1984,West1994,Escobar1995}. These methods also have extensions to grouped data and the estimation of a set of related density functions, the problem considered here.  Important methods for grouped data include the analysis of densities model of \cite{Escobar1999} which uses a DPM model for each density in which the base measure for the mixing densities is the same and given a DPM prior; the hierarchical Dirichlet process \citep{Teh2006} where mixing distributions are given DP priors with a common base measure which is given a DP prior; Dirichlet process mixture of ANOVA models \citep{DeIorio2004} where the atoms in a Dirichlet process 
are modelled as dependent on a covariate following an ANOVA type dependence structure; and the hierarchical model of \cite{Muller2004} in which distributions are modelled as a mixture of a common and group specific component, with these components being given DP mixture priors.  

One alternative to DP based methods in Bayesian nonparametrics is the use of Gaussian processes \citep{Leonard1978,Thorburn1986,Lenk1988, Lenk1991,Tokdar2007a, Tokdar2007b}.  However, to the best of our knowledge hierarchical versions of Gaussian process priors for grouped data situations have not been developed in the literature.  Gaussian process methods can be attractive because the parameters in the resulting priors allow very easy expression of relevant prior information, such as smoothness of the densities and, in the hierarchical setting, the extent of sharing information between groups.  For the DP mixture based methods, on the other hand, generally prior information must be expressed through a prior on a mixing distribution, through which it is difficult to adequately express similar prior beliefs.  One reason that Gaussian process density estimation methods are not more popular is perhaps the computational difficulty. Such difficulties are even more acute in the hierarchical setting involving 
grouped data.  

Here we introduce a hierarchical Gaussian process prior, a reformulated version of the prior proposed by \cite{Adams2009}, and discuss tractable methods for computation with this prior.  Our construction, besides being able to handle an arbitrary number of hierarchy levels, provides a convenient way of expressing prior beliefs regarding both the degree of similarity between the densities and the nature of their characterising features, such as smoothness and support. 
We also establish a remarkably different approach for making inference. Instead of relying on Markov chain Monte Carlo (MCMC) methods to draw samples from the posterior distribution, as is commonly implemented in other approaches to Gaussian process density estimation \citep{Adams2009} which can suffer from poor performance, we alternatively introduce an approximate Bayesian computation (ABC) \citep{Beaumont2002} functional regression-adjustment to draw approximate samples from the posterior. The use of ABC to estimate functional (infinite-dimensional) objects in itself represents an important contribution to the ABC literature.   
 
In Section \ref{sec:prior}, we introduce the hierarchical Gaussian process prior and present an algorithm for sampling  data from this prior. Section \ref{sec:inference}  describes the inferential strategy for estimating the infinite-dimensional parameter. Performance of the proposed density estimator is investigated in a simulation study in Section \ref{sec:simulation}. Finally, in Section \ref{sec:application}, we use the proposed estimator to compare rural high school exam performance across states in Brazil.

\section{The hierarchical Gaussian process prior}
\label{sec:prior}

We specify a hierarchical Gaussian process prior on the set of densities $f_i(x)$, $i=1,\dots,g$,  as follows:
\begin{subequations}
  \label{eq:prior}
    \begin{align}
     \label{eq:map}  f_i(x) & = \frac{L(Z_i(x)) b(x | \mathbf{\phi})}{c_i(\phi, Z_i(x))}  \\
     \label{eq:GP} Z_i(x) & \sim \mathcal{GP}(\mu(x), k(x, x' | \theta_1)) \\
     \mu(x) & \sim \mathcal{GP}(m(x), k(x, x' | \theta_2)) \\
     \label{eq:Hprior} (\theta_1, \theta_2) & \sim \pi(\cdot).
  \end{align}
\end{subequations}
Here $L(z)=1 / (1 + \exp(-z))$ denotes the logistic function, $b(x | \phi)$ is an arbitrarily chosen parametric base density with hyperparameters $\phi$, and $\mathcal{GP}(\mu(x), k(x, x'|\theta))$ represents a Gaussian process with mean function $\mu(x)$ and covariance function $k(x, x'|\theta)$ with parameters $\theta$. The function $m(x)$ is a conveniently chosen function discussed further below, $\pi(\cdot)$ is a hyperprior for the parameters of each covariance function, and $c_i(\phi, Z_i(x))=\int L(Z_i(x)) b(x | \phi) dx$ is the normalising constant of $f_i(x)$.  Gaussian processes are a common tool for modelling functional data and can be understood as an infinite-dimensional generalisation of the Gaussian distribution. A comprehensive review of Gaussian processes is found in \cite{GPbook}.  See also \cite{Shi2011} for a book length treatment of Gaussian processes in functional data analysis.

Equation (\ref{eq:map}) defines a deterministic map from auxiliary functions $Z_i(x)$ to the normalised densities $f_i(x)$. Operating on the transformed space avoids difficulties otherwise implied by the intrinsic properties of density functions -- namely, that they be non-negative and integrate to one.
Under this prior, all marginal densities are considered potentially related, and bound by a common Gaussian process mean function $\mu(x)$ (Equation \ref{eq:GP}), which is itself unknown. The above formulation,  though simple, will be shown to be highly adaptable, and capable of adequately describing a wide range of possible prior beliefs. For ease of presentation only two levels of hierarchy are described above, but it is straightforward to incorporate multiple hierarchical levels, as is illustrated in the analysis of Brazilian school exam performance in Section \ref{sec:application}.

For this article, we adopt the squared exponential kernel, $k(x, x') = \sigma^2 \exp(-\alpha(x - x')^2)$ as covariance function, although other options are easily accommodated. The hyperparameters $(\theta_1, \theta_2) = (\sigma_1, \alpha_1, \sigma_2, \alpha_2)$ play a crucial role in the behaviour of $f_i(x)$. Figure 1 shows $g=5$ densities drawn from the proposed prior under various parameter settings for $(\sigma_1, \alpha_1, \sigma_2, \alpha_2)$. The standard uniform distribution was chosen as the base density $b(x|\phi)$ and we set $m(x)=-10$ (the motivation for this choice will be made clear in Section \ref{sec:inference}). 
\begin{figure}[htb]
	\centering
	\subfloat[\label{fig:case1}]{\includegraphics[width=5cm,height=5cm,angle=-90]{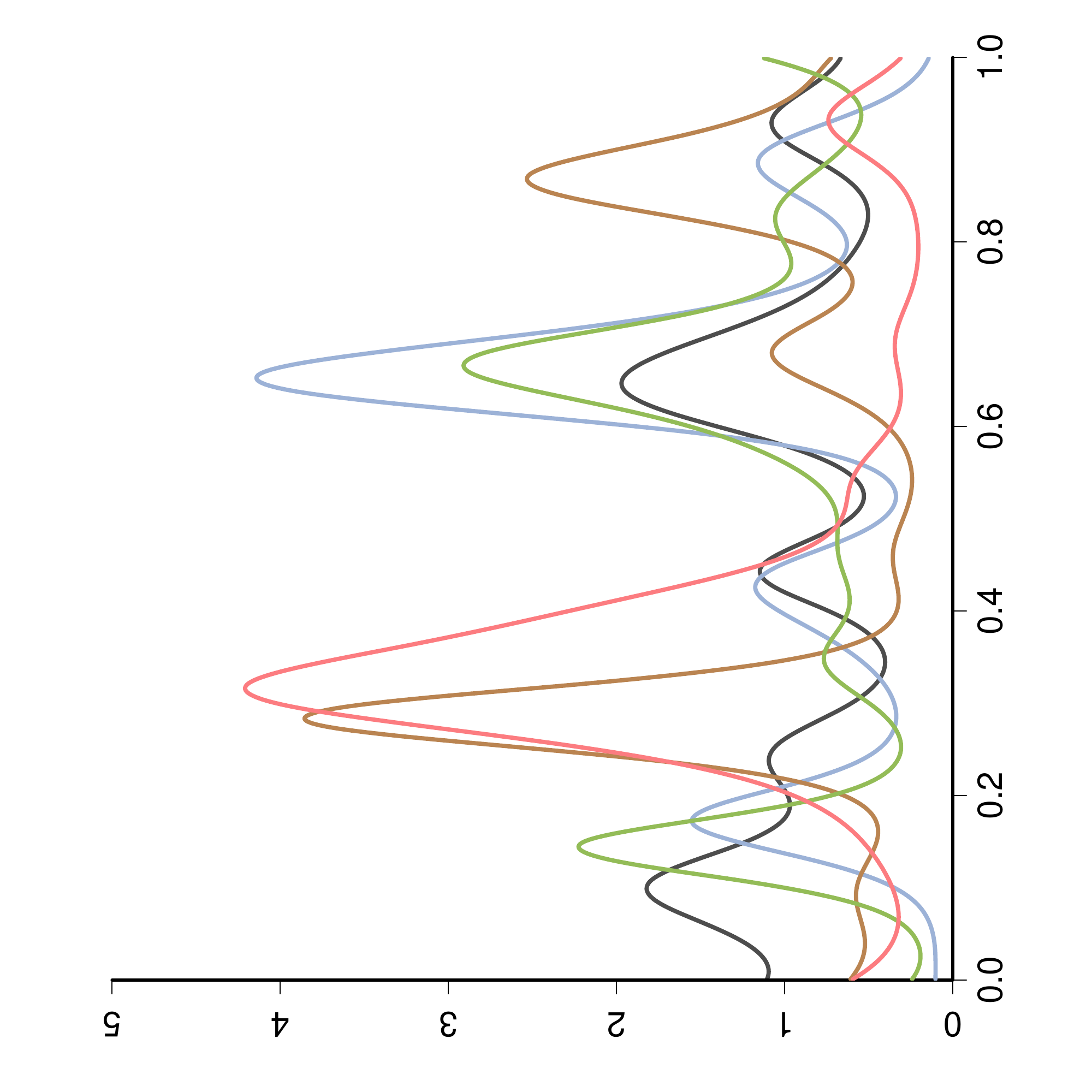}}
	\subfloat[\label{fig:case2}]{\includegraphics[width=5cm,height=5cm,angle=-90]{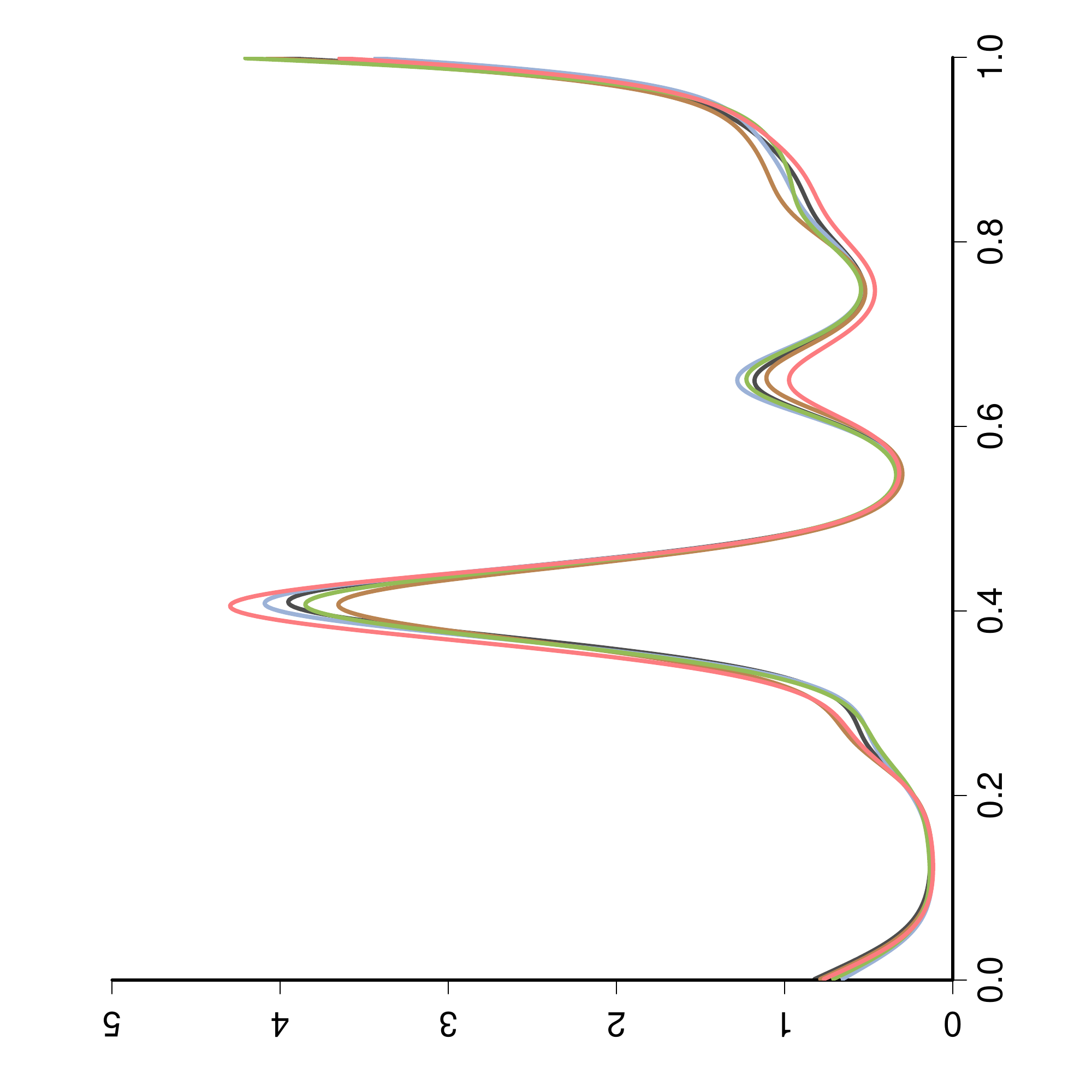}}
	\\
	\subfloat[\label{fig:case3}]{\includegraphics[width=5cm,height=5cm,angle=-90]{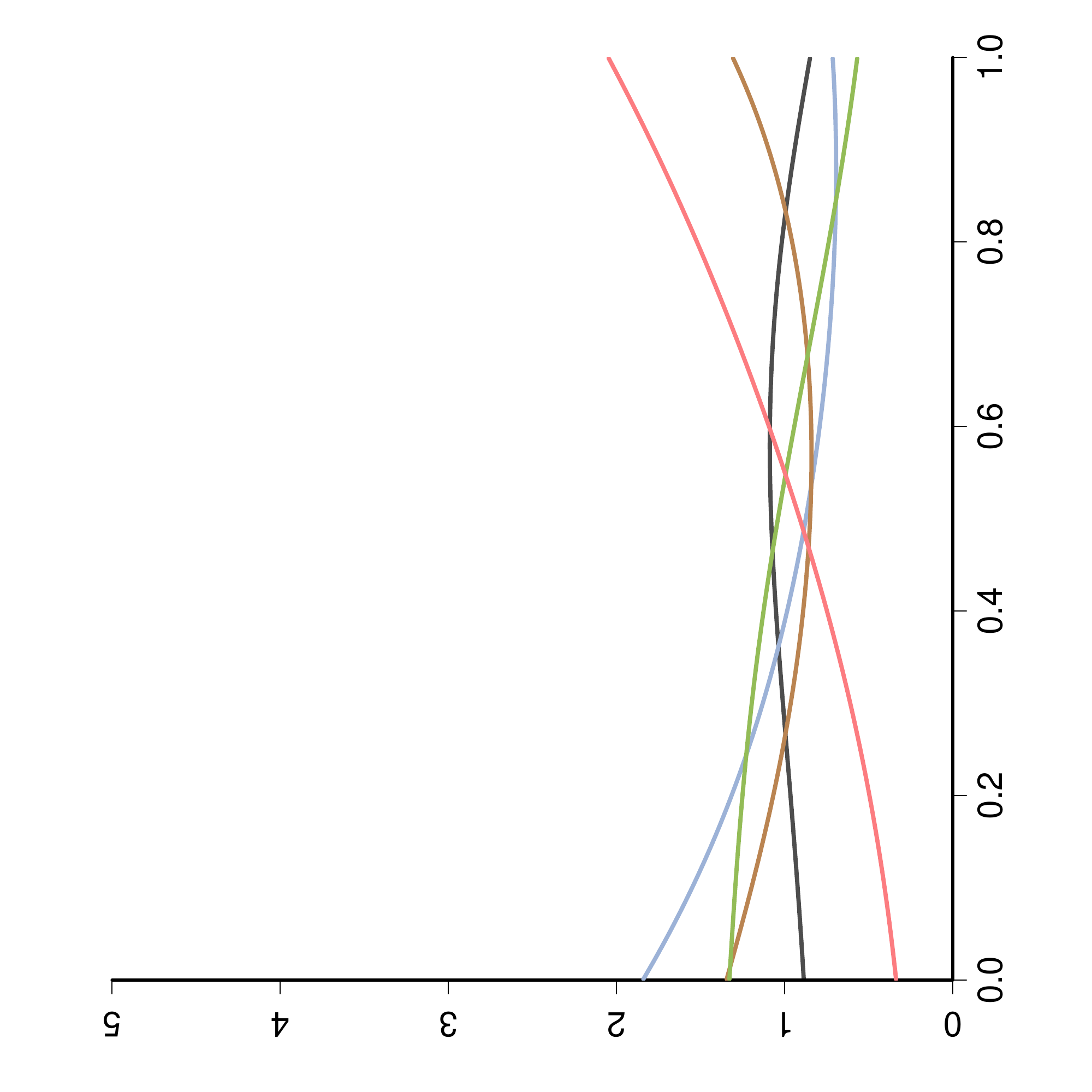}}
	\subfloat[\label{fig:case4}]{\includegraphics[width=5cm,height=5cm,angle=-90]{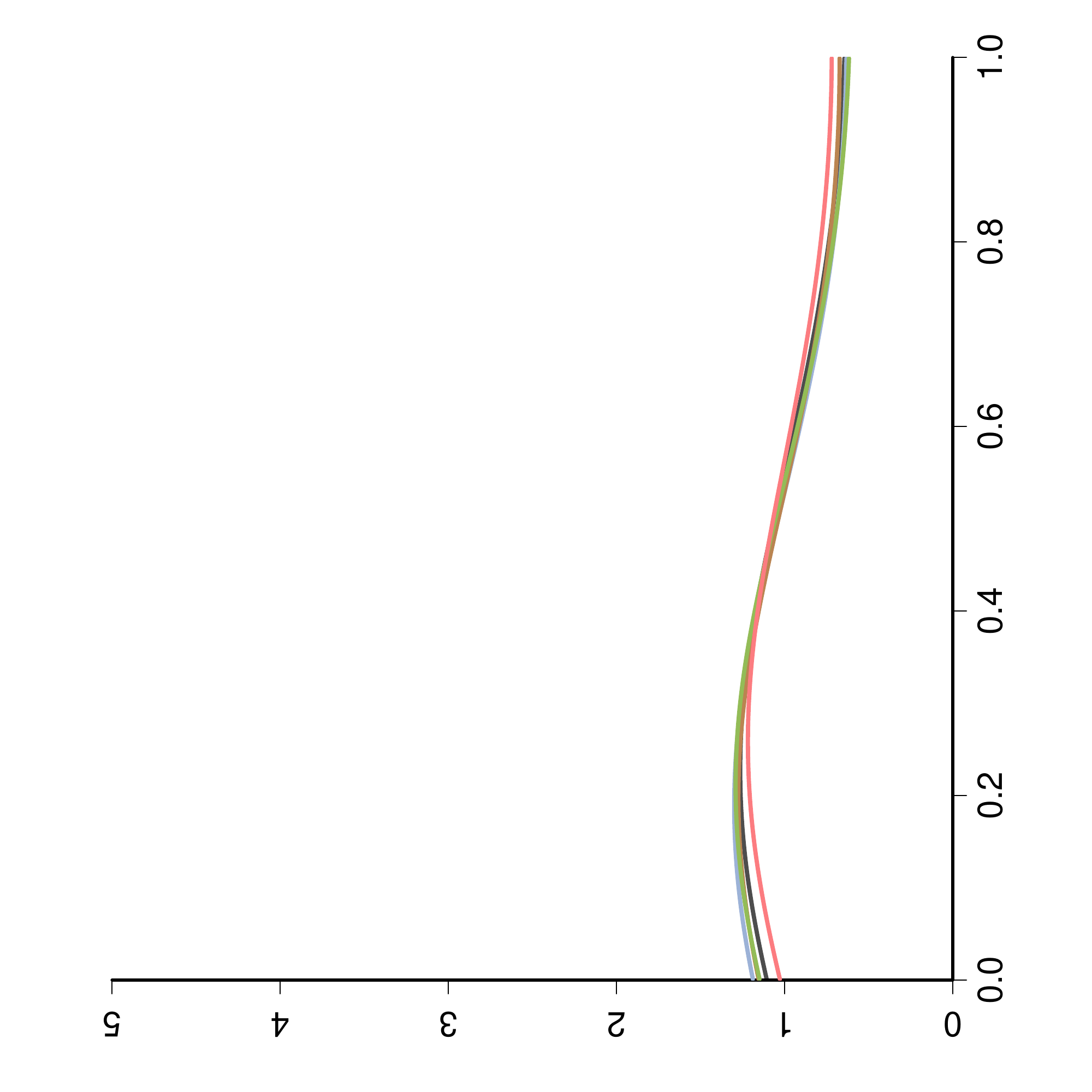}}
	\caption{\small Samples from the prior distribution based on the squared exponential function with $g=5$, under varying prior conditions. The above panels show: (a) independence and dissimilarity to the base density ($\sigma_1 = 1, \alpha_1 = 100, \sigma_2 = 0.1, \alpha_2 = 100$); (b) dependence and dissimilarity to the base density ($\sigma_1=0.1, \alpha_1=100, \sigma_2=1, \alpha_2=100$); (c) independence and similarity to the base density ($\sigma_1=1, \alpha_1=1, \sigma_2=0.1, \alpha_2=1$), and (d) dependence and similarity to the base density ($\sigma_1=0.1, \alpha_1=1, \sigma_2=1, \alpha_2=1$).}
	\label{fig:prior.samples}
\end{figure}

While the $\alpha$'s are length-scale parameters that determine the speed with which the covariance decays as a function of the distance $|x - x'|$, the $\sigma$ parameters dictate how similar the densities are to $b(x | \phi)$, with larger $\sigma$ indicating less similarity. The ratio $\sigma_1 / \sigma_2$ defines how closely related the different densities are: if the ratio is large, the densities are nearly independent (Figures \ref{fig:case1} and \ref{fig:case3}). On the other hand, if the ratio is small, most of the variability is due to the variation of the common mean, which results in very similar marginal densities (Figures \ref{fig:case2} and \ref{fig:case4}).

\subsection{Sampling from the prior}
\label{subsec:sampling}

Approximate draws from this prior may be naturally obtained by sequentially sampling the unknown parameters from the top (Equation (\ref{eq:Hprior})) to the bottom (Equation (\ref{eq:map})) of the hierarchy. It is possible to derive an exact sampler using a variation of Algorithm 3.1 in \cite{Adams2009}. However, for a faster simulator, we favour a finite-dimensional approximation as in \cite{Tokdar2007a}, which results in the so-called surrogate prior.

In particular, we first sample the hyperparameters $(\theta_1, \theta_2)\sim\pi(\cdot)$ from  (\ref{eq:Hprior}). Next, define $x_{low}$ and $x_{high}$ to be the lower and upper $\beta$-quantiles of $b(x | \phi)$, where $\beta$ is chosen to be small, and construct a regular grid of $k$ points on the range $[x_{low}, x_{high}]$, i.e., $\psi_1 < \ldots < \psi_k$, with $\psi_j = x_{low} + (x_{high} - x_{low}) (j - 1)  / (k - 1)$, for $j=1, \ldots, k$.

It follows from the Gaussian process definition that $\mu^\psi = (\mu(\psi_1), \ldots, \mu(\psi_k))^\top$ follows a $k$-variate normal distribution with mean $m^\psi = (m(\psi_1), \ldots, m(\psi_k))^\top$ and covariance matrix $\Sigma$, with typical element $\Sigma_{ij} = k(\psi_i, \psi_j | \theta_2)$. Sampling $\mu^\psi$ is therefore trivial. Similarly, we obtain $Z_i^{\psi}=(Z_i(\psi_1),\dots, Z_i(\psi_k))^\top$, $i=1,\ldots, g$,
as draws from the multivariate normal distribution
$N(\mu^\psi,\Lambda)$, with covariance matrix defined as $\Lambda_{ij}=k(\psi_i,\psi_j|\theta_1)$. The normalising constants $c_i(\phi, Z_i(x))$ may then be estimated using numerical integration. 
Through (\ref{eq:map}), this uniquely determines the corresponding densities $f_1^\psi, \ldots, f_g^\psi$.
 
Each vector $Z_i^\psi$ and $f_i^\psi$ represents a discretised observation of the corresponding underlying function. Continuous approximations for $Z_i(x)$ and $f_i(x)$ can then be obtained by fitting B-splines to each of these vectors via least squares, resulting in $\tilde{Z}_i(x)$ and $\tilde{f}_i(x)$, respectively. The quality of this approximation is controlled by the number of points in the grid, $k$, and the number of basis functions used.
Finally, data $X_{ij}$ may be generated from $\tilde{f}_i(x)$ using a rejection sampling algorithm with $b(x | \phi)$ as the proposal distribution. A candidate $x^*$ is accepted with probability $\tilde{f}_i(x^*) / [M b(x^* | \phi)]$, where $M = \max\{\tilde{f}_i(x) / b(x | \phi)\}$. The value of $M$ is unknown, but may be determined by numerical search.

\section{An approximate Bayesian inferential procedure}
\label{sec:inference}

The resulting posterior distribution is given as
\begin{eqnarray*}
    p(\mathbf{Z}, \theta | \mathcal{D}) & = & \frac{p(\mathbf{Z}, \theta) p(\mathcal{D} | \mathbf{Z})} {\int \int p(\mathbf{Z'}, \theta') p(\mathcal{D} | \mathbf{Z'}) d\mathbf{Z'} d\theta'} \\
    & \propto & \pi(\theta_1, \theta_2) p(\mu(x) | \theta_2) \prod_{i=1}^{g} p(Z_i(x) | \mu(x), \theta_1) \prod_{j=1}^{n_i} \frac{L(Z_i(x_{ij})) b(x_{ij} | \phi)} {c_i(\phi, Z_i(x_{ij}))},
\end{eqnarray*}
where $p$ denotes a joint or conditional distribution, $\theta=(\theta_1,\theta_2)$, $\mathbf{Z}$ compactly denotes the set of functions $Z_1(x), \ldots, Z_g(x)$ and $\mathcal{D}$ represents the observed data. The above posterior distribution is computationally difficult to work with directly.
As an alternative we develop an approximate Bayesian computation procedure. ABC methods have been extensively developed to draw samples from an approximation to the posterior distribution, based only on the ability to sample data  from the model, without the need to evaluate the posterior directly. See e.g. \cite{Beaumont2002,blum+nps13,nott+fms14} for more details on ABC methods.
As part of this inference, we extend the ideas behind the ABC regression-adjustment \citep{Beaumont2002} to the functional (infinite-dimensional) setting.

Samples from the posterior distribution can be used to estimate any quantity of interest, such as the mean or quantiles of $f_i(x)$. Here, the posterior mean $\mathbb{E}[f_i(x) | \mathcal{D}]$, which is a function itself, is a Bayes estimator for $f_i(x)$. Note that while any given sample from $p(Z_i(x) | \mathcal{D})$ corresponds to a unique sample from $p(f_i(x) | \mathcal{D})$, the reverse does not hold.

\subsection{ABC for Gaussian process density estimation}%
\label{subsec:naive}

Approximate Bayesian computation methods implement an approximate Bayesian inference based on the matching of simulated and observed summary statistics. A typical rejection-sampling ABC algorithm applied to this particular context would be as follows: 
\begin{enumerate}
\item Draw a sample (a set of density functions) from the hierarchical Gaussian process prior using the method described in Section \ref{subsec:sampling}, and generate synthetic data from these densities of the same size as the observed data. 

\item Summarise the synthetic data using a set of summary statistics. We adopt the well known kernel density estimator, computed for each group, $i=1,\ldots,g$. 
\item If the resulting summary statistic is, by an appropriate metric, similar to the summary statistic computed over the observed data, accept the densities generated in step 1.  Otherwise these are rejected. 
\item Repeat steps 1--3 to produce $m$ accepted samples.
\end{enumerate}
The popularity and computational simplicity of the kernel density estimator makes it a suitable candidate to play the role of the summary statistic in the ABC procedure. In particular, we employ a Gaussian kernel and set the bandwidth as $h=\hat{\sigma}(4/3n)^{-1/5}$ where $\hat{\sigma}$ denotes the sample standard deviation \citep{Scott1992}. We denote by $K_{i\ell}(\psi_j)$ the resulting kernel density estimate obtained from the data in group $i$ in synthetic dataset $\ell$ evaluated at $\psi_j$, for $i=1,\ldots,g$, $\ell=1,\ldots,m$. Alternative density estimators can be adopted, providing they are computationally fast.

Step 3 above requires the specification of a measure of similarity between the synthetic and observed data. Here, a natural and convenient option is to define the divergence of synthetic dataset $\ell$ from the observed data as 
\begin{equation}
   \label{eq:KL} D_\ell = \sum_{i=1}^{g} \sum_{j=1}^{k} \left| \log (K^{obs}_{i}(\psi_j)) - \log (K_{i\ell}(\psi_j)) \right| \, K^{obs}_{i}(\psi_j),
\end{equation}
where $|\cdot|$ denotes the absolute value operator, and where $K^{obs}_{i}(\psi_j)$  denotes the kernel density estimator described above applied to the observed data in group $i$, evaluated at $\psi_j$. 

This divergence measure is closely related to the sum of the Kullback-Leibler divergences of the kernel density estimates computed over the observed and synthetic data.

In order to produce more accurate posterior samples, we follow  \cite{Beaumont2002} and weight the samples according to how well the synthetic data reproduces the observed data through (\ref{eq:KL}) using the Epanechnikov kernel
$w(D) =  1 - (D / \delta)^2$ for $D \leq \delta$ and $0$ otherwise.  
Here $\delta$ is a threshold which determines how close the synthetic and observed summary statistics must be before the sample densities are accepted as an approximate  draw from the posterior distribution in step 3. Its value is often determined in terms of a quantile of the

\subsection{A functional regression-adjustment}%
\label{subsec:adjustment}

As with all ABC algorithms, the approximation error associated with the ABC method described in Section \ref{subsec:naive} can be substantial, unless the threshold $\delta$ is small. However, if $\delta$ is too small, then the acceptance rate of the algorithm will be prohibitively low. \cite{Beaumont2002} proposed the regression-adjustment as a simple approach to adjust the posterior samples obtained with $\delta>0$, so that $\delta\approx 0$.
In the simple finite-dimensional, parametric model considered, a (local) linear relationship
\[
	\vartheta=\alpha+\beta^\top(s-s_{obs})+\varepsilon
\]
is assumed to hold between the model parameters $\vartheta$ and the vector of simulated and observed summary statistics, $s$ and $s_{obs}$, where $\varepsilon$ denotes a zero mean error. If this model holds, then
\[
	\vartheta^*=\vartheta-\hat{\beta}^\top(s-s_{obs}),
\]
where $\hat{\beta}$ is the least-squares estimate of $\beta$, is an approximate draw from the posterior with $s=s_{obs}$ (i.e. $\delta=0$.) See \citet{Beaumont2002,blum+f10,blum+nps13} for further discussion of regression-adjustment methods.
In the present setting, we may similarly perform a regression-adjustment,  but using a functional regression model \citep{FDA2005} due to the functional nature of the parameters and summary statistics.

From the definition of the logistic function,  $\log(L(z))\approx z$, for negative and sufficiently small $z$. Accordingly, Equation (\ref{eq:map}) can be rewritten as
\begin{align}
     \label{eq:Truerelation} Z_i(x) \approx \log \left(\frac{f_i(x) c_i(\phi, Z_i(x))}{b(x | \mathbf{\phi})}\right).
\end{align}
This approximation is accurate up to the fourth decimal place for $Z_i(x) \leq -10, \forall x$, which can be easily achieved through appropriate choice of $m(x)$. In this article we set $m(x)=-10$. A useful, additional advantage of using a negative mean function for $\mu(x)$ is that this prior does not give strong support to unrealistic, flat-peaked densities, which can happen if one sets $m(x)=0$.
Equation (\ref{eq:Truerelation}) forms the basis to appropriately specify a functional regression model. For $i=1, \ldots, g$, we write
\begin{align}
     \label{eq:reg.mod} \tilde{Z}_{i\ell}(x) = \text{offset} + \gamma^i_0(x) + \gamma^i_1(x) \log(\tilde{K}_{i\ell}(x)) + \gamma^i_2(x) \overline{\log}(\tilde{K}_{-i,\ell}(x)) + \epsilon^\ell(x),
\end{align}
where $\tilde{Z}_{i\ell}(x)$ denotes a B-spline fitted to $Z_{i\ell}(\psi_1),\ldots,Z_{i\ell}(\psi_k)$, 
$\gamma^i_0(x), \gamma^i_1(x)$ and $\gamma^i_2(x)$ are functional regression parameters to be estimated, and $\epsilon^\ell(x)$ is a zero-mean error term. The offset term is given by $\log \left( \frac{b(x | \mathbf{\phi})} {c_i(\phi, \tilde{Z}_{i\ell}(x))}\right)$ and is computed through (\ref{eq:map}), $\tilde{K}_{i\ell}(x)$ denotes a B-spline fitted to $K_{i\ell}(\psi_1),\ldots,K_{i\ell}(\psi_k)$, and 
\begin{equation}
	\label{eq:logKtilde}
	\overline{\log}(\tilde{K}_{-i,\ell}(x))= \frac{1}{g-1} \sum_{\substack{1 \leq h \leq g \\ h \neq i}} \log(\tilde{K}_{h\ell}(x)). 
\end{equation}
With this construction, while $\log(\tilde{K}_{i\ell}(x))$ conveys information from the \textit{i}-th group itself, $\overline{\log}(\tilde{K}_{-i,\ell}(x))$ explores what has been observed in the remaining groups. Accordingly, $\gamma^i_2(x)$ defines how much (and where) strength should be borrowed.

From (\ref{eq:Truerelation})--(\ref{eq:logKtilde}) it can be seen that the data relates to $\tilde{Z}_{i}(x)$ through the function
\begin{align*}
\log\bigg([\tilde{K}_{i}(x)]^{\gamma_1(x)} \prod_{\substack{1 \leq h \leq g \\ h \neq i}}[\tilde{K}_{h}(x)]^{\gamma_2(x) / (g-1)}\bigg).
\end{align*}
This means we are ultimately using the logarithm of a weighted geometric mean of the kernel densities to model the dependent functional object. By no means we are suggesting, though, that our final estimate is a weighted geometric average of the kernel densities -- there are other terms in the model. The weights, which are estimated from (\ref{eq:reg.mod}) via least squares, are subject to the condition that all groups other than group \textit{i} have the same weight $\gamma_2(x) / (g-1)$. This enforced condition simplifies the problem without losing too much information.

Based on (\ref{eq:reg.mod}), approximate samples from the marginal posterior distribution $p(Z_i(x) | \mathcal{D})$ can then be obtained by performing the functional regression adjustment
\begin{equation}
	\label{eqn:funcregadj}
     \tilde{Z}^*_{i\ell}(x) = \tilde{Z}_{i\ell}(x) - \hat{\gamma}^i_1(x) [\log(\tilde{K}_{i\ell}(x)) - \log(\tilde{K}^{obs}_{i}(x))] - \hat{\gamma}^i_2(x) [\overline{\log}(\tilde{K}_{-i,\ell}(x)) -
     \overline{\log}(\tilde{K}^{obs}_{-i}(x))]
\end{equation}
where $\hat{\gamma}^i_1(x)$  and $\hat{\gamma}^i_2(x)$ are least-squares estimates of $\gamma^i_1(x)$  and $\gamma^i_2(x)$ (obtained using the {\tt R} package {\tt fda}, \cite{fda}), and where $\tilde{K}^{obs}_{i}(x)$ and $\tilde{K}^{obs}_{-i}(x)$ are the equivalent summary statistics  obtained from the observed data. 

Note that the right-hand side of the proposed functional regression model (\ref{eq:reg.mod}) is only defined for strictly positive kernel estimates. If necessary, this can be artificially satisfied by adding an arbitrary tiny constant to the kernel estimates. This modified summary statistic ensures the model is well-defined and may provide extra stability.

Finally, while the above regression models are intuitive and seem sensible, as with the parametric regression-adjustment \citep{Beaumont2002} they remain only a model for the perhaps complex relationships that exist between the density estimates and the summary statistics. Given that the regression model is fitted to those samples for which $D_\ell<\delta$, this means that the local-linearity assumption of (\ref{eq:reg.mod}) may be adequate for many situations providing that $\delta$ is sufficiently small. As for the parametric regression-adjustment, more complex models can be developed if required \citep{blum+f10}.

\section{ A simulated example}
\label{sec:simulation}

In this section we present a simulated analysis to illustrate some important features of the ABC density estimator. We construct $g=10$ groups, with sample sizes $5, 20, 35, \ldots, 140$. The base density is taken to be the standard uniform distribution and the priors for the parameters of the squared exponential covariance functions are
$\sigma_1,\sigma_2\sim\mbox{Gamma}(3,5)$ and $\alpha_1,\alpha_2\sim\mbox{Gamma}(1,0.1)$, which are fairly uninformative. We generate $50,000$ samples from the prior distribution and accept the $m = 5,000$ samples closest to the observed data. We use $k=100$ grid points and fit the B-splines functions using $50$ basis terms.

\begin{figure}[htb]
  \centering
  \subfloat[\footnotesize{True densities} \label{fig:true}]{\includegraphics[width=8cm,height=8cm,angle=-90]{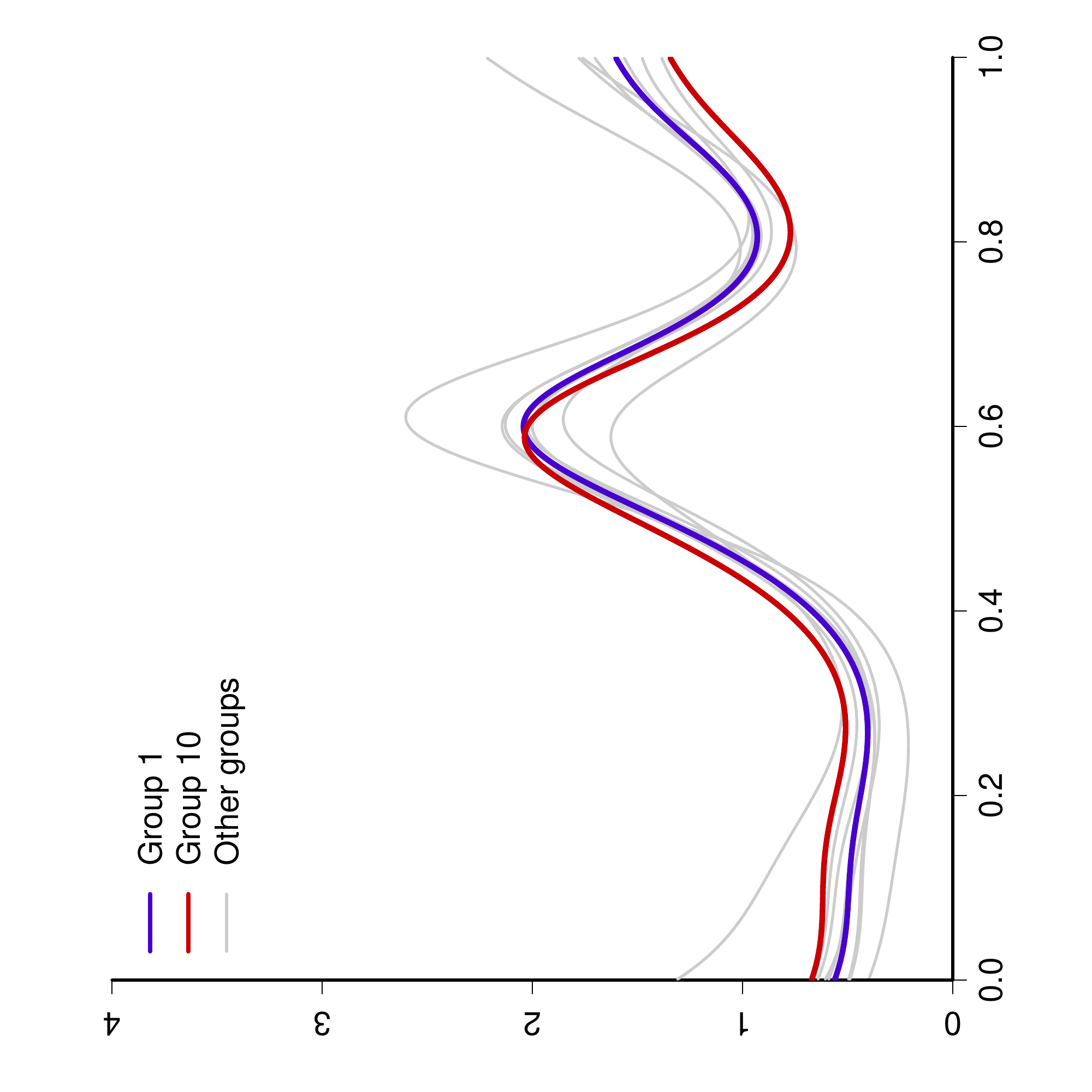}}
  \subfloat[\footnotesize{Kernel density estimates} \label{fig:kernels}]{\includegraphics[width=8cm,height=8cm,angle=-90]{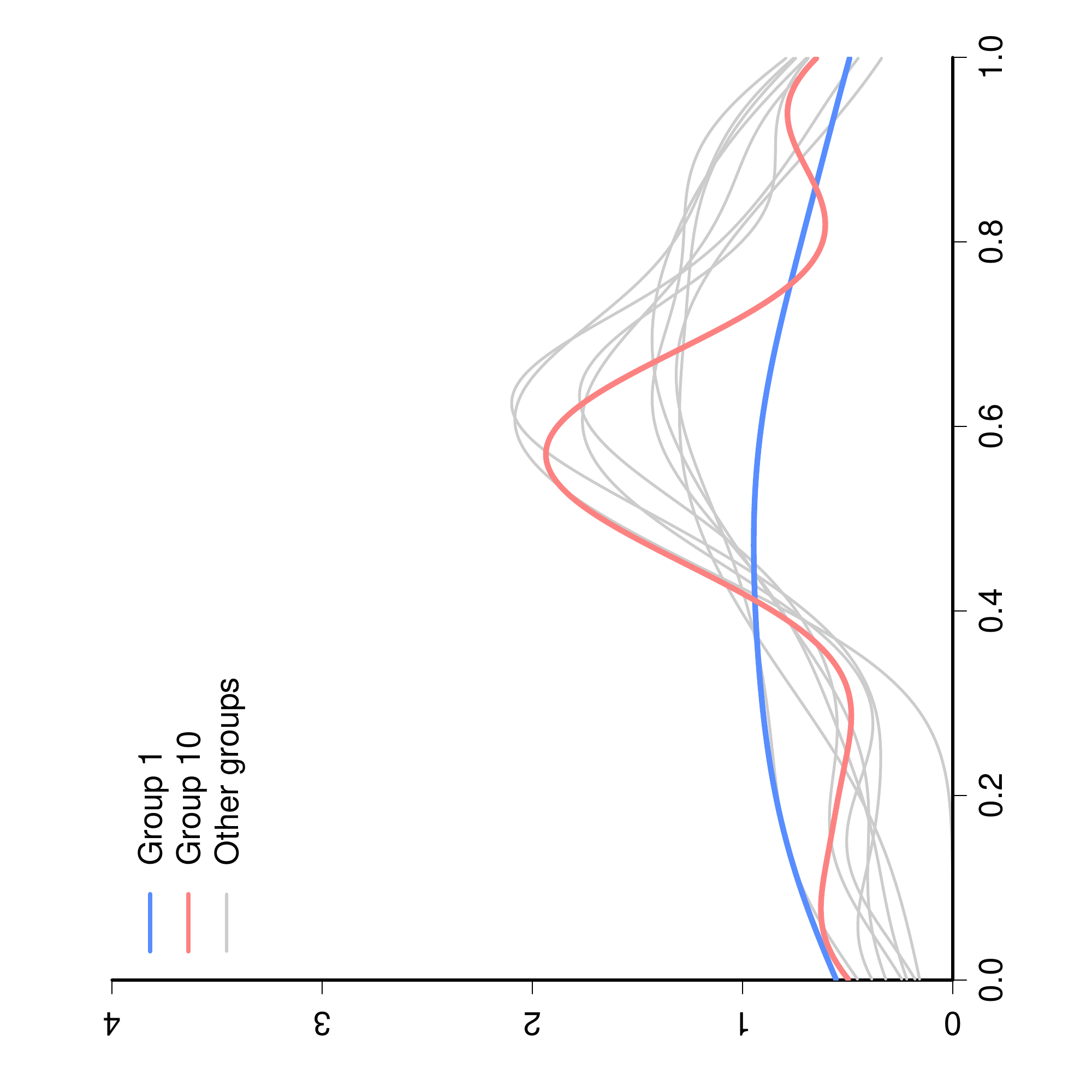}}
  \caption{\small (a) True  densities and (b) kernel density estimates of the simulated data from $g=10$ groups. Group 1 (5 datapoints) and group 10 (140 datapoints) are highlighted.}
  \label{fig:simulated1}
\end{figure}
\begin{figure}[hbt]
  \centering
  \subfloat[\footnotesize{Estimated regression coefficients} \label{fig:coefs}]{\includegraphics[width=8cm,height=8cm,angle=-90]{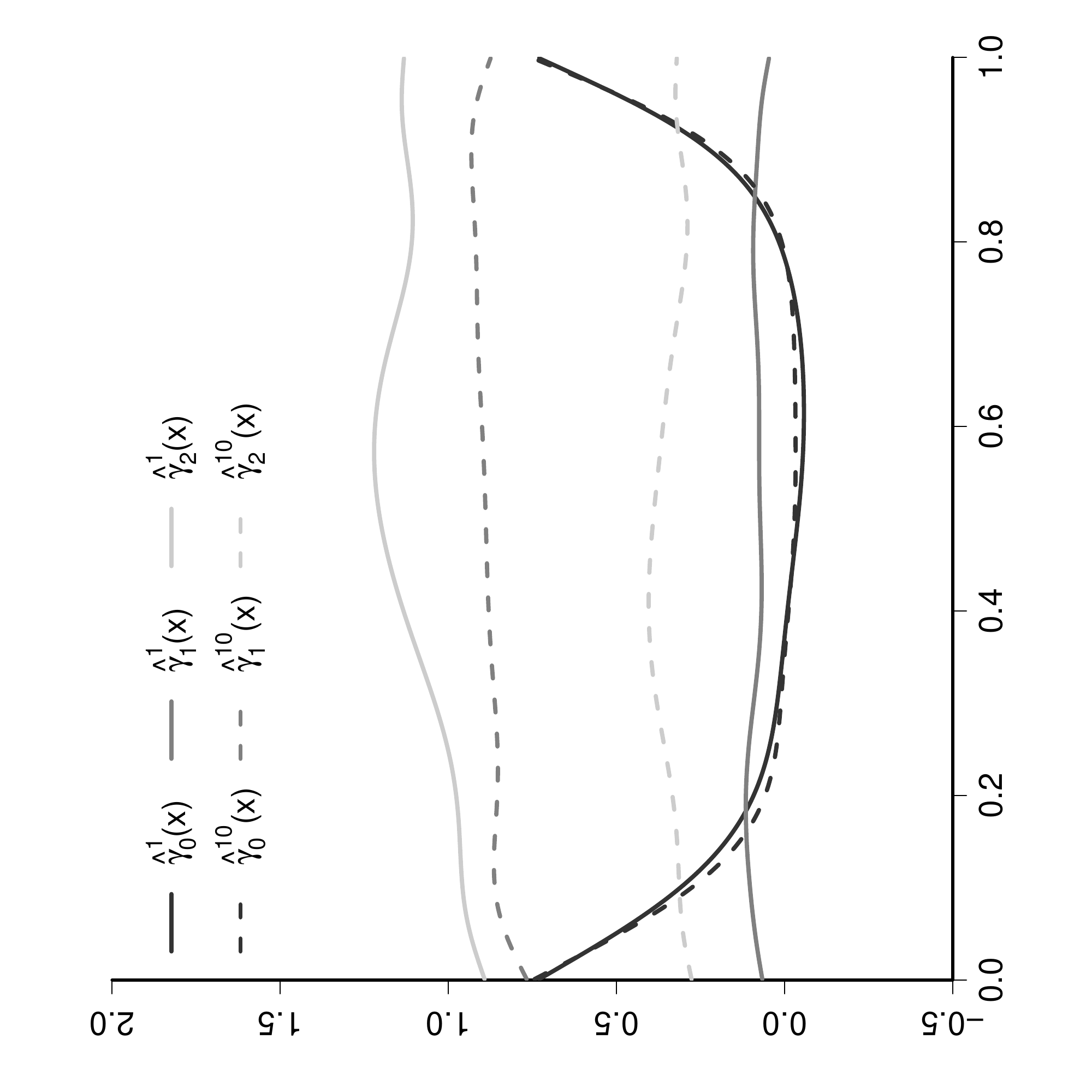}}
  \subfloat[\footnotesize{Residual plot} \label{fig:residuals}]{\includegraphics[width=8cm,height=8cm,angle=-90]{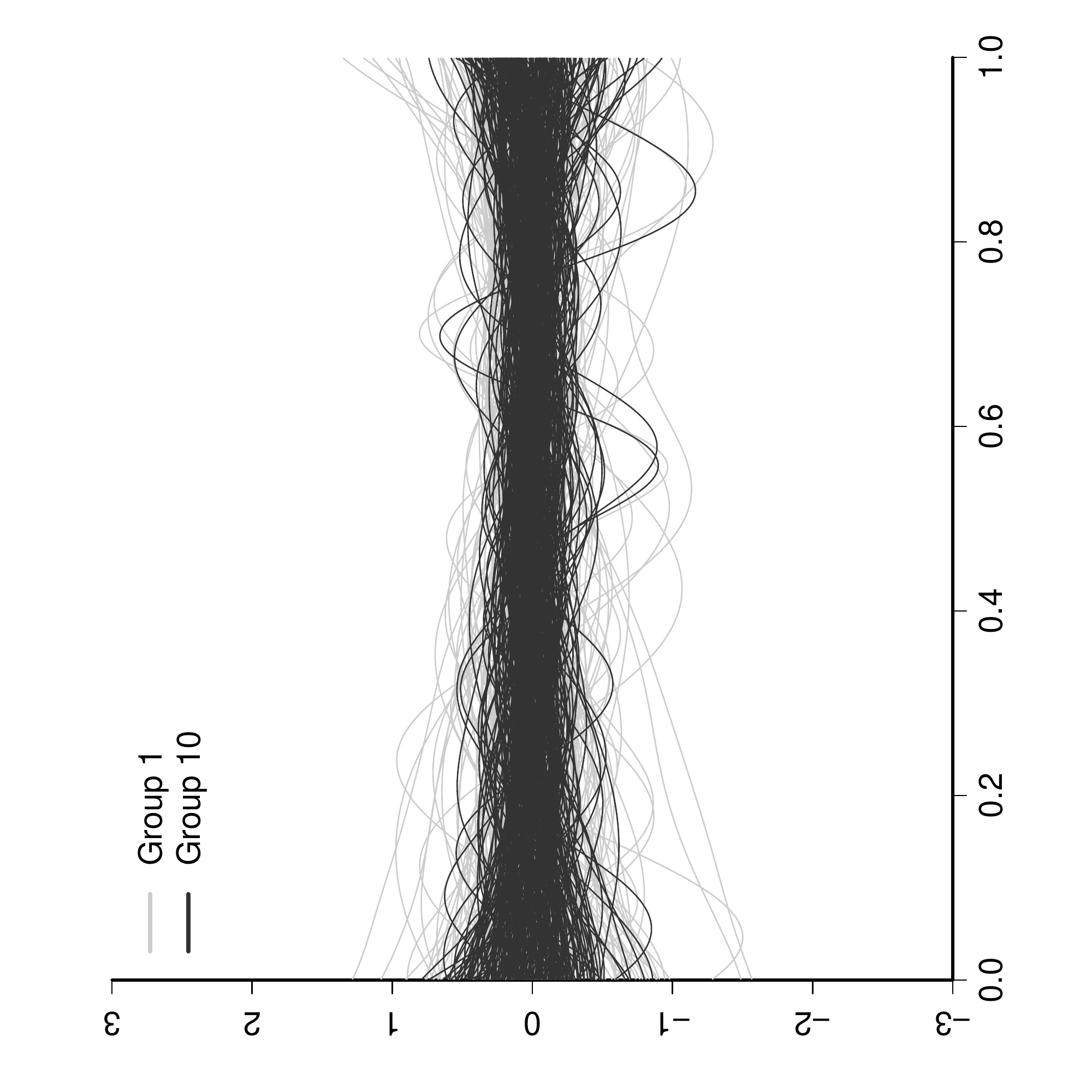}}
  \caption{\small (a) Least-squares estimates of the functional regression coefficients $\gamma_0(x), \gamma_1(x)$ and $\gamma_2(x)$ for groups 1 (solid lines) and 10 (dashed lines); (b) Least-squares functional residuals for groups 1 (black lines) and 10 (grey lines).}
  \label{fig:simulated2}
\end{figure}

We focus our attention on groups 1 and 10, as they have the smallest and largest sample sizes, respectively. 
Figure \ref{fig:simulated1} shows the true densities (left panel) and kernel density estimates (right panel) of each of the groups, with the latter also representing a first ``crude" estimation of the respective densities. Although the density estimate for group 10 looks reasonable, the estimation for group 1 is clearly unsatisfactory due to the low sample size. The well known boundary-bias of this kernel density estimator \citep{Wand1991,Chen1999,Jones1993,Jones2007,Dai2010,Geenens2014} is clearly visible as a downward dip in the density estimates near the borders of the support.

Figure \ref{fig:simulated2} (left panel) shows the estimated functional regression coefficients for groups 1 (solid lines) and 10 (dashed lines). The intercepts $\gamma^1_0$ and $\gamma^{10}_0$ (black lines) are effectively zero except for upward ticks near the boundaries. In effect, the regression model is both identifying and attempting to correct for the boundary-bias in the density estimates. 
The other regression parameters describe the contribution of  $\log(\tilde{K}^{obs}_{i}(x))$ and $\overline{\log}(\tilde{K}^{obs}_{-i}(x))$ in estimating $\tilde{Z}_i(x)$. For Group 1, virtually all relevant information comes from the kernel estimates of the other groups, as $\hat{\gamma}_1^1$ (solid grey line) is near zero over the entire range. For Group 10, the situation is reversed -- most of the information comes from its own kernel density ($\hat{\gamma}_1^{10}(x)$ is large), although it still borrows strength from other groups as $\hat{\gamma}_2^{10}(x)$ (light grey dashed line) is far from zero everywhere. All parameter estimates are not constant, indicating that the relationship varies 
in different regions of the data-space. For groups 2 to 9 (plots not shown),  a steady transition is observed between the results of groups 1 and 10.

The right panel of Figure \ref{fig:simulated2} shows the plot of the estimated functional residuals given by
$\hat{\varepsilon}_{i\ell} = \tilde{Z}_{i\ell}(x) - \hat{Z}_{i\ell}(x)$, for $i=1,10$, where $\hat{Z}_{i\ell}(x)$ denotes the fitted regression model. For both groups, the residuals are centred on zero everywhere, with greater variability near the boundary as expected. In addition, the residuals are more variable for group 1 than for group 10 due to the differences in the sample sizes.

\begin{figure}[htp]
  \centering
  \subfloat[\footnotesize{Samples from the posterior - Group 1}\label{fig:posterior1}]{\includegraphics[width=8cm,height=8cm,angle=-90]{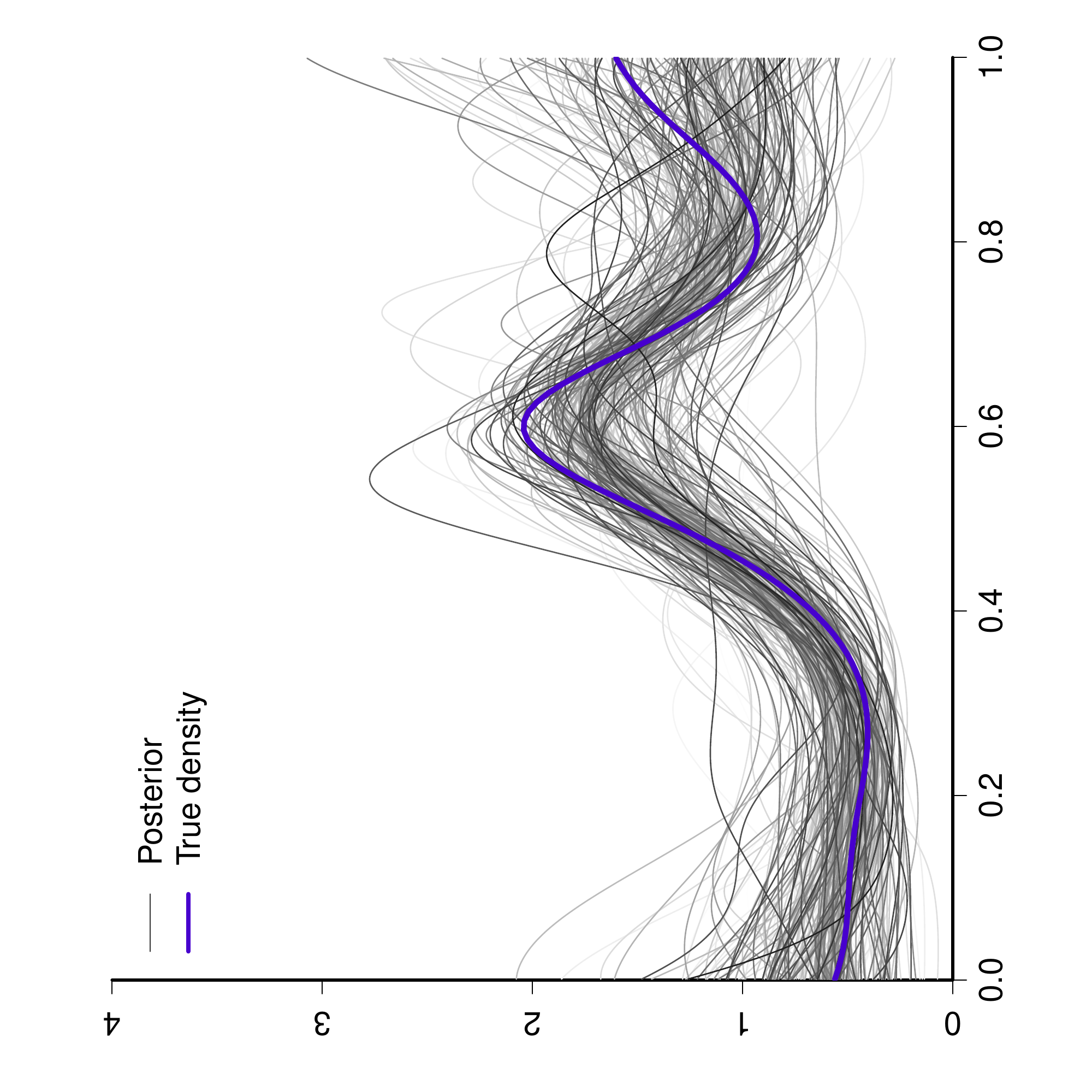}}
  \subfloat[\footnotesize{$95\%$ credibility intervals - Group 1}\label{fig:summary1}]{\includegraphics[width=8cm,height=8cm,angle=-90]{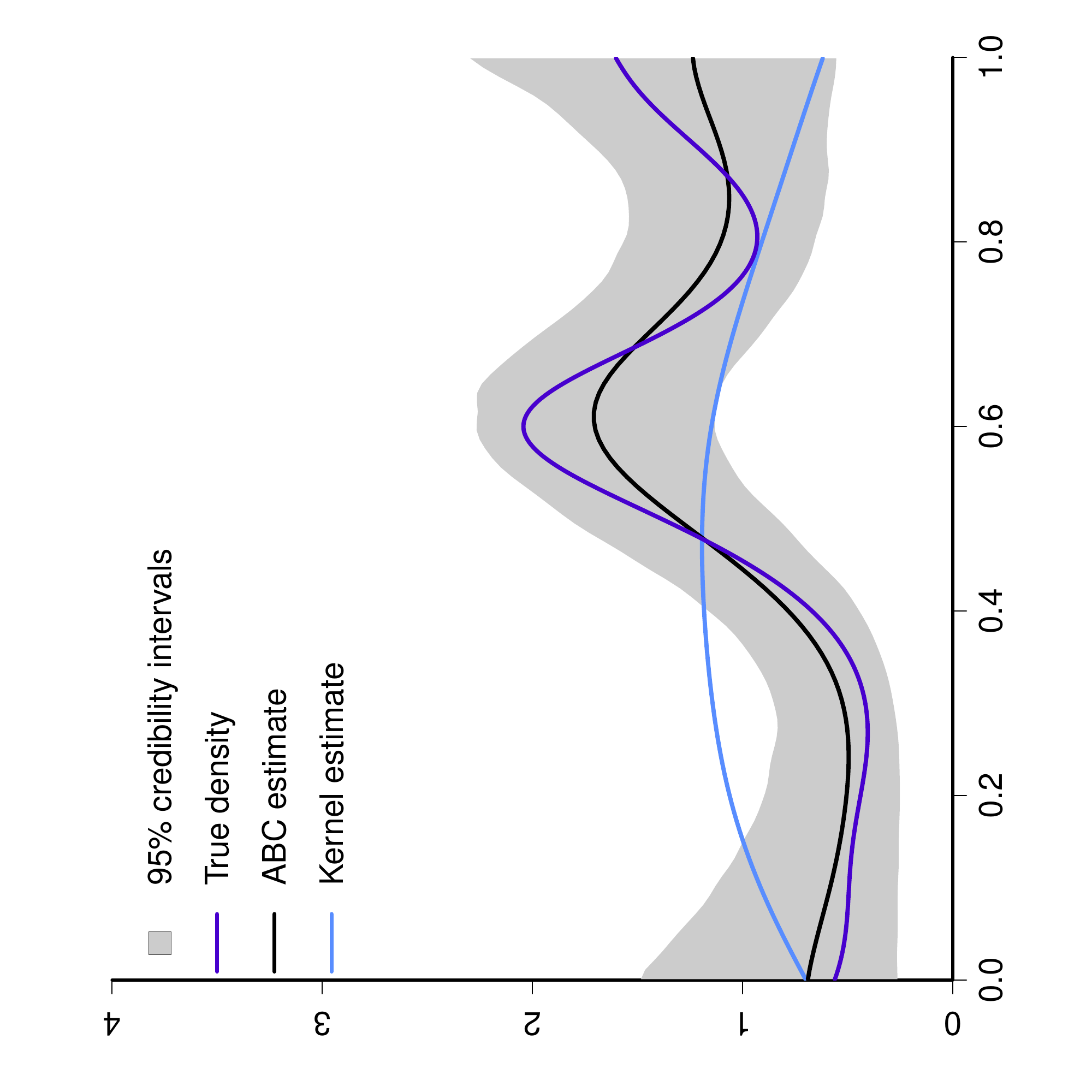}}
  \\
  \subfloat[\footnotesize{Samples from the posterior - Group 10}\label{fig:posterior2}]{\includegraphics[width=8cm,height=8cm,angle=-90]{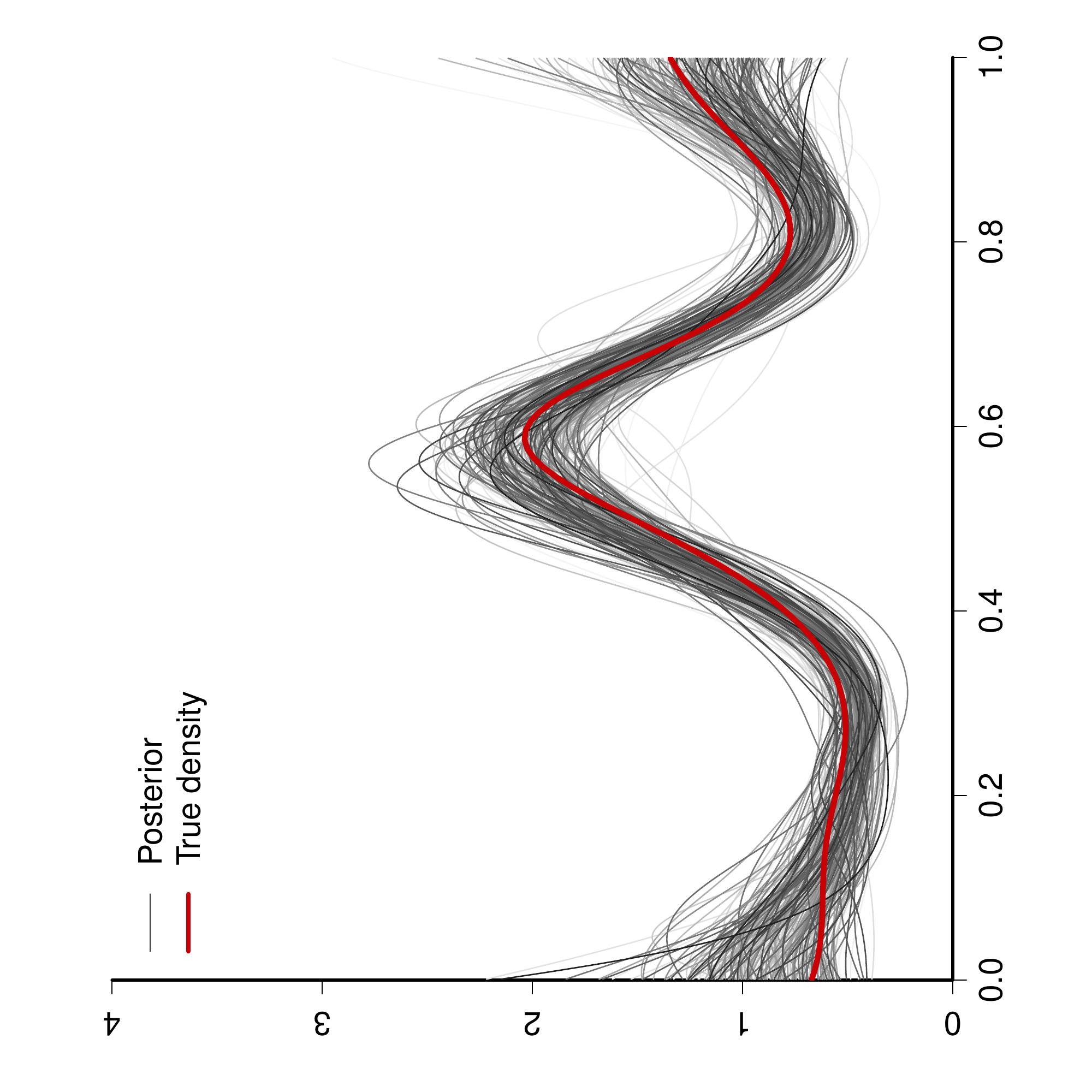}}
  \subfloat[\footnotesize{$95\%$ credibility intervals - Group 10}\label{fig:summary2}]{\includegraphics[width=8cm,height=8cm,angle=-90]{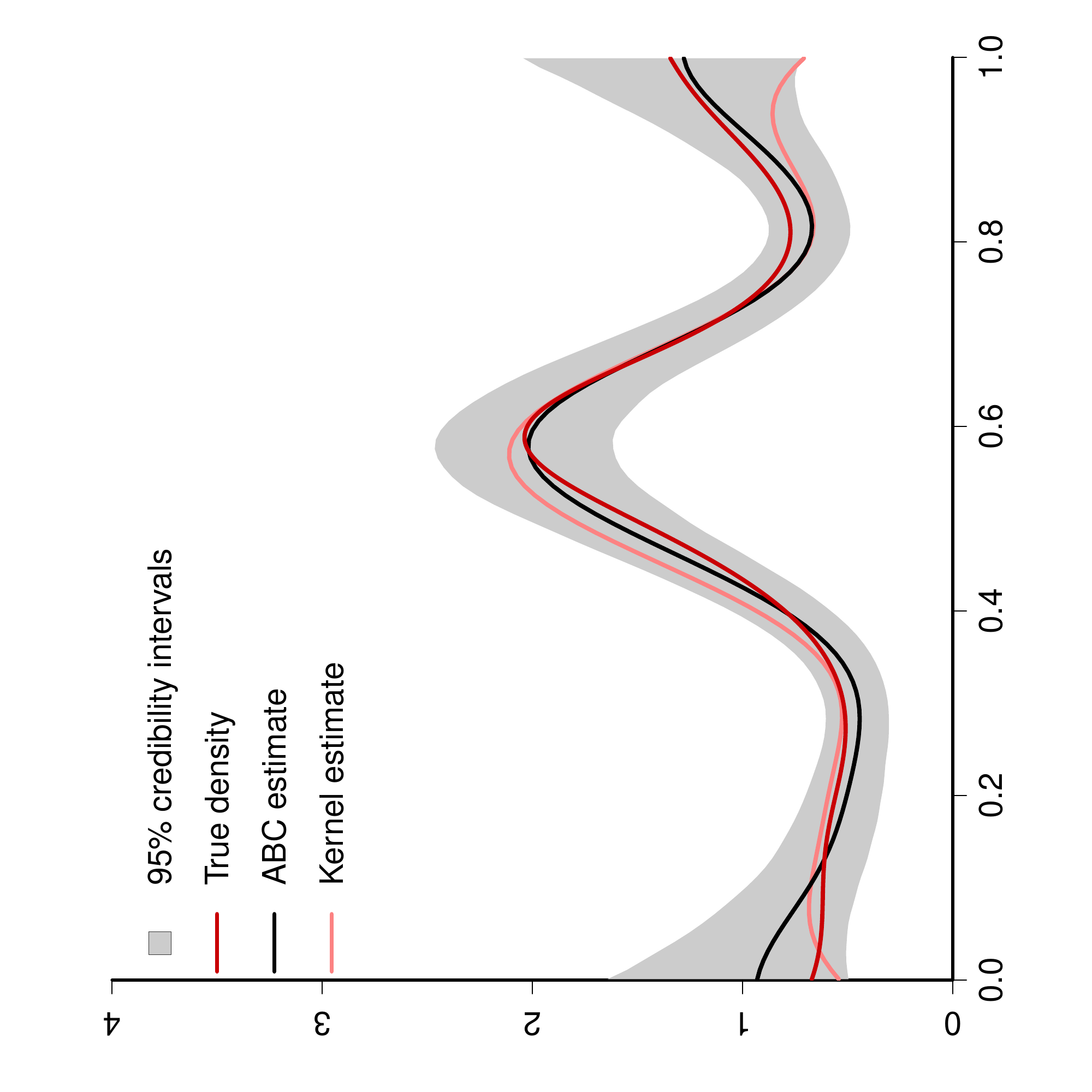}}
  \caption{\small Left panels: Samples from the posterior distribution for groups 1 and 10 (grey lines), with a darker line indicating greater sample weight. True density is indicated by the coloured line. Right panels: Comparison of the true densities, the initial kernel density estimate and the ABC posterior mean $\mathbb{E}[f_i(x)|\mathcal D]$ with pointwise 95\% central credible intervals, for groups 1 and 10.}
  \label{fig:simulated3}
\end{figure}

The left panels in Figure \ref{fig:simulated3} illustrate 250 randomly chosen approximate samples from the posterior. The darker lines indicate larger sample weights. It is clear that the samples cluster around the true densities, with the variability for group 1 (small sample size) being greater than that of group 10.
The right panels in Figure \ref{fig:simulated3} depict pointwise 95\% central credibility intervals of the posterior distribution, the posterior mean $\mathbb{E}[f_i(x)|\mathcal D]$, the initial group kernel density estimate and the true density function. As expected, the ABC posterior mean strongly outperforms the initial kernel density estimate. For group 1, the hierarchical sharing of information between density estimates has produced a substantially improved and more accurate estimate. Even for group 10, which has the most data, the posterior mean is more accurate, and less biased at the boundaries than the initial kernel density estimate.

\section{An analysis of high school exam performance in Brazil}
\label{sec:application}

Appropriate assessment of school performance is paramount to efficiently manage large educational systems. In 1998 the Brazilian government introduced the high school national exam, Enem (\textit{Exame Nacional do Ensino Medio}), which annually evaluates high school students in all states of the country.  
Enem is used by the Department of Education to set the education agenda and strategically define public policies, and  by Federal universities and other educational institutes as part of their student selection criteria. 
According to the Organisation for Economic Co-operation and Development,  
students who attend schools in urban areas tend to perform at higher levels than other students \citep{Pisa2013}. They conclude that beyond socio-economic reasons, schools in urban areas are generally larger and benefit from better resources and greater autonomy.

   \begin{figure}[htp]
  \centering
  \subfloat[\label{fig:Ext.case1}]{\includegraphics[width=6cm,height=6cm,angle=-90]{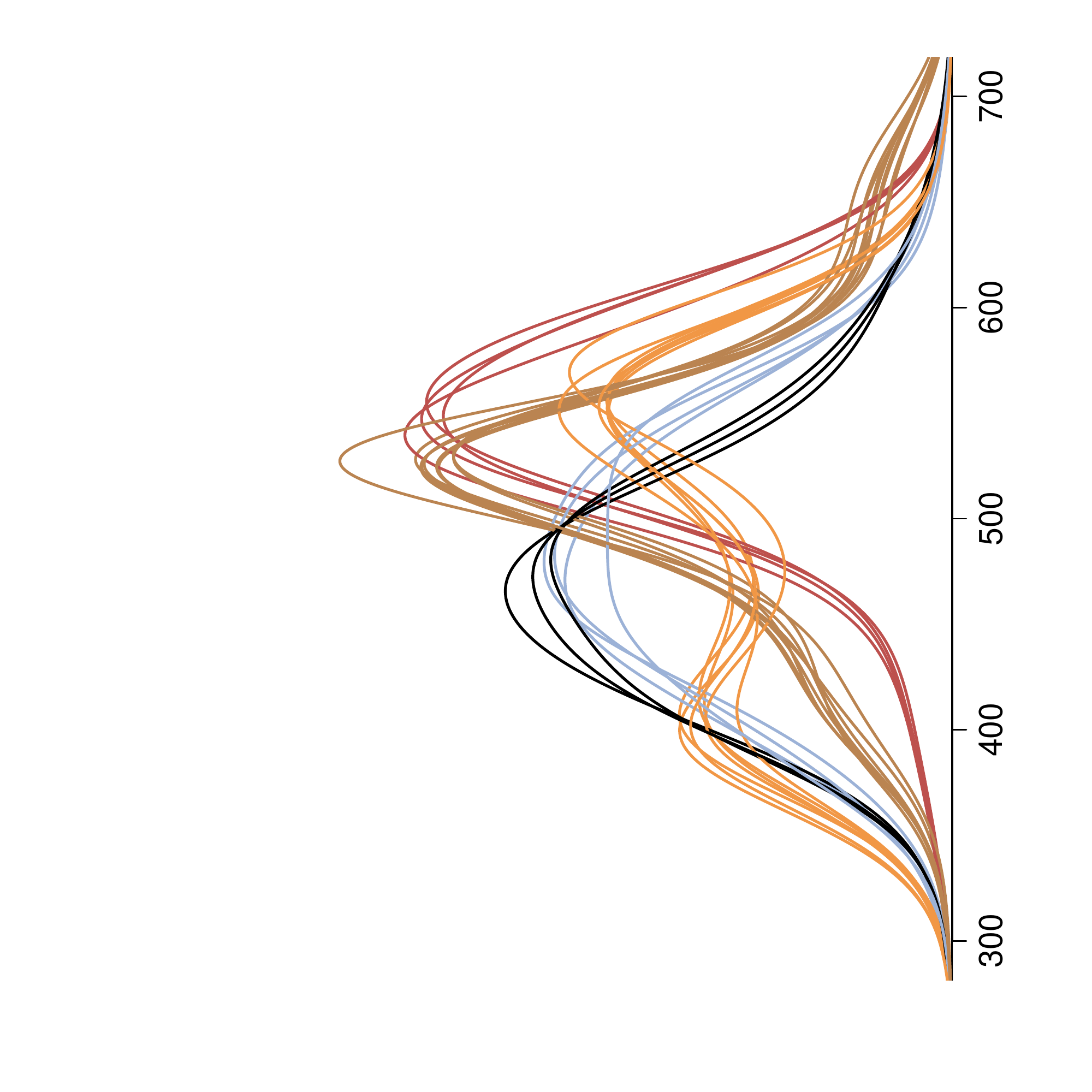}}
  \subfloat[\label{fig:Ext.case2}]{\includegraphics[width=6cm,height=6cm,angle=-90]{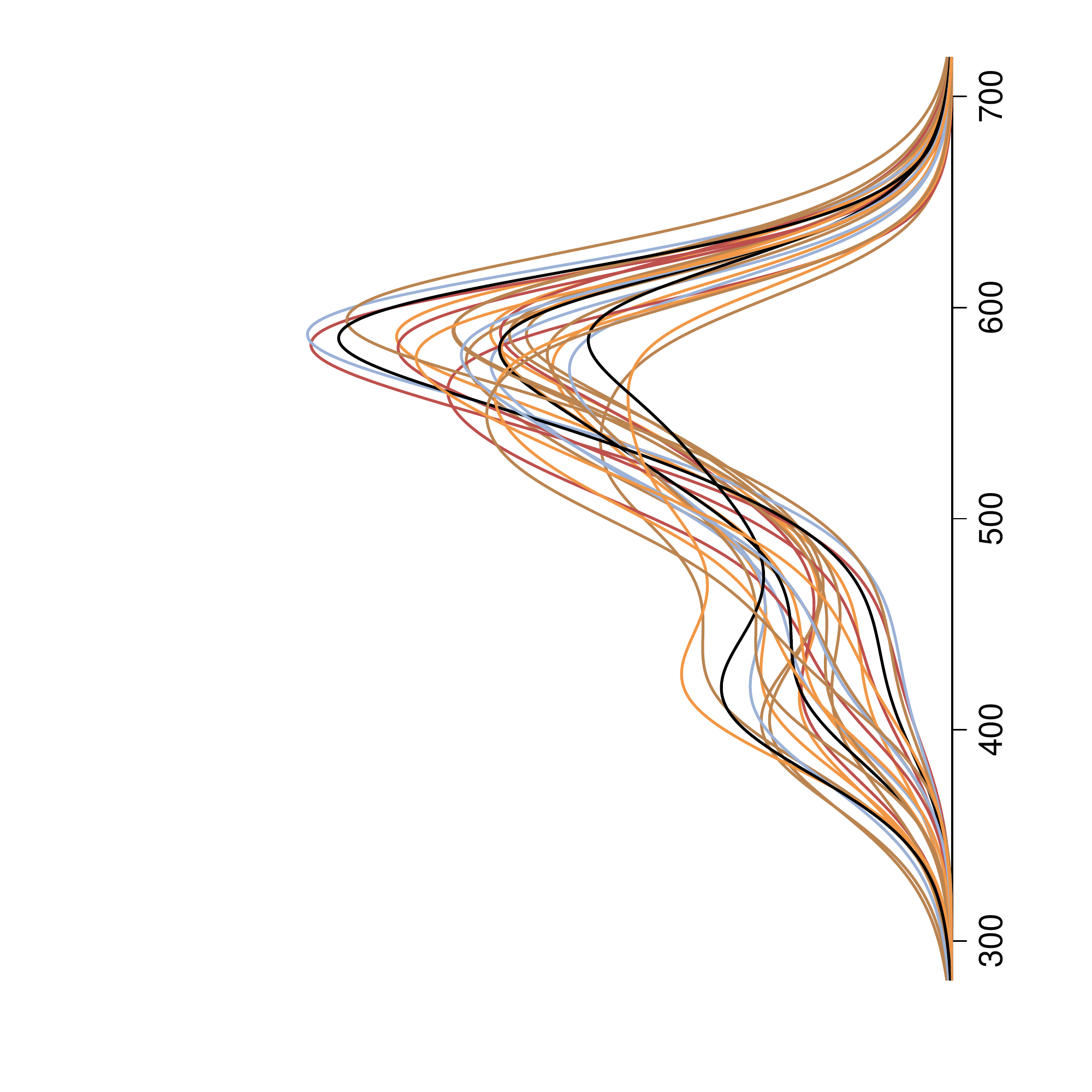}}
  \\
  \subfloat[\label{fig:Ext.case3}]{\includegraphics[width=6cm,height=6cm,angle=-90]{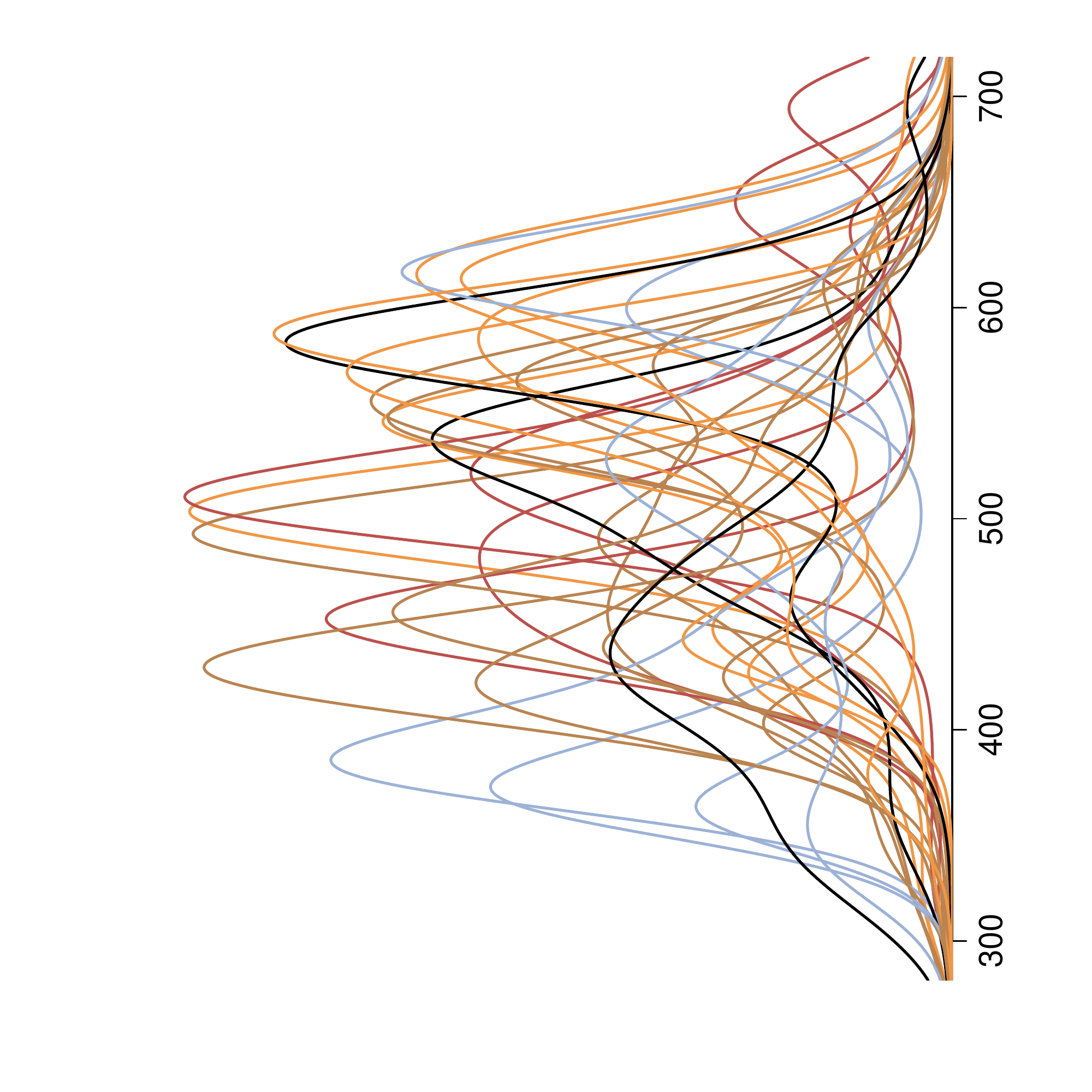}}
  \subfloat[\label{fig:Ext.case4}]{\includegraphics[width=6cm,height=6cm,angle=-90]{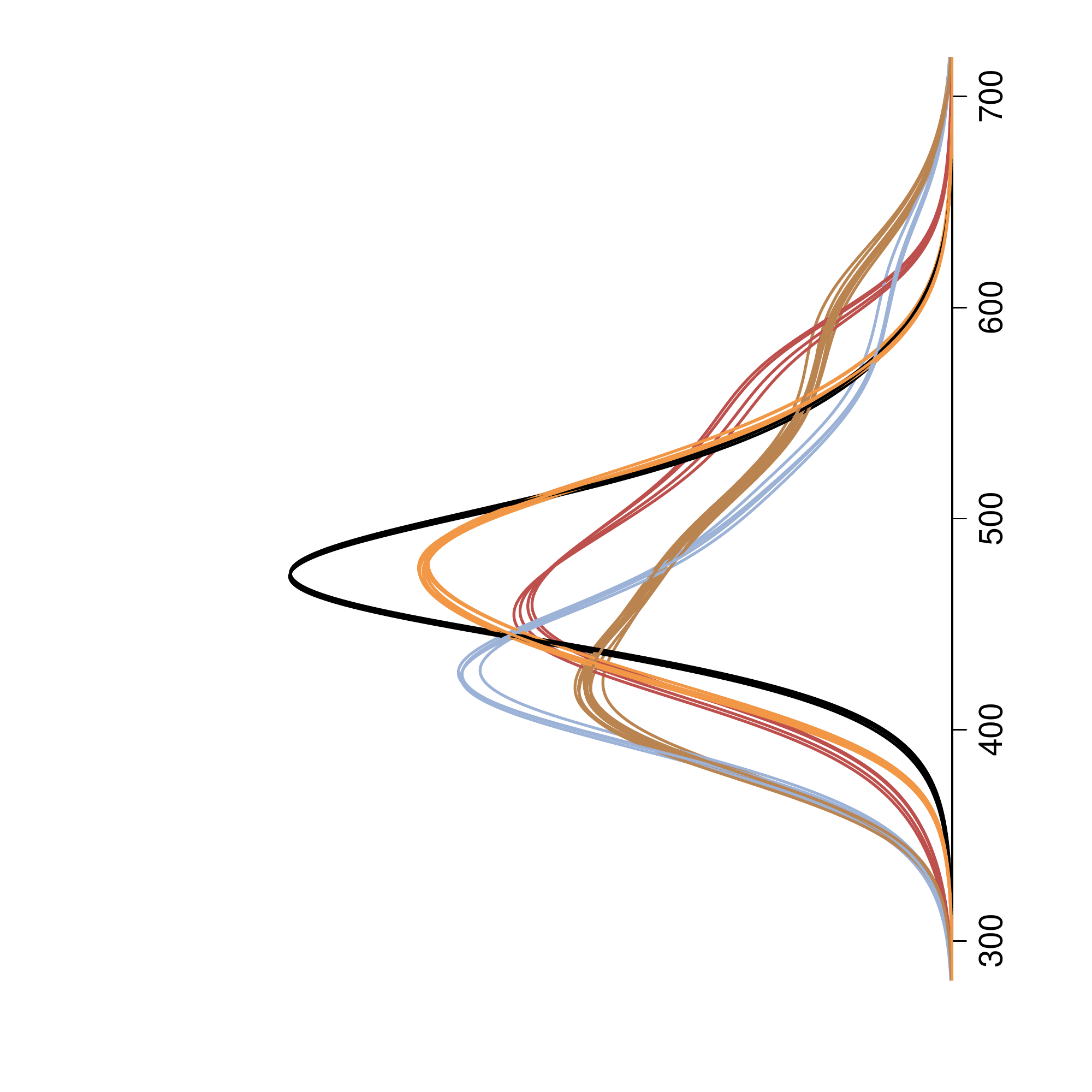}}
  \caption{Samples from the two-level prior distribution based on the squared exponential covariance function with $g_1=27$ state groups and $g_2=5$ regional groups under varying parameter conditions. The above panels show: (a) moderate state and regional dissimilarity; (b) moderate state dissimilarity and regional similarity; (c) strong state and regional dissimilarity, and (d) strong state similarity and regional dissimilarity. Density colours indicate regional membership.}
  \label{fig:application1}
\end{figure}

We analyse data extracted from the 2012 Enem dataset to evaluate
rural school performance. The observational units are the rural schools themselves, each of which is measured by the mean grade of its students.
Interest is in estimating and comparing the grade densities of each of the 27 states.
 Only schools with more than 10 student attending the exam are considered. The number of recorded rural schools in each state varies from 3 in Roraima, to 101 in Sao Paulo.
The data are available from \verb+http://portal.inep.gov.br/basica-levantamentos-acessar+.

The 27 states are divided into 5 regions -- North, Northeast, West Central, Southeast and South. We expect that there may be some degree of similarity between states in the same region, and so a reasonable prior should allow us to express this formally. For this purpose, we can extend the prior described in (\ref{eq:map})--(\ref{eq:Hprior}) to include an additional level. More precisely,
\begin{eqnarray*}
     f_i(x) & = & \frac{L(Z^1_i(x)) b(x | \mathbf{\phi})}{c_i(\phi, Z_i^1(x))}  \\
     Z^1_i(x) & \sim & \mathcal{GP}(Z^2_{\delta(i)}(x), k(x, x' | \sigma_1, \alpha_1)) \\ 
     Z^2_j(x) & \sim & \mathcal{GP}(Z^3(x), k(x, x' | \sigma_2, \alpha_2)) \\ 
     Z^3(x) & \sim & \mathcal{GP}(m(x), k(x, x' | \sigma_3, \alpha_3)), 
\end{eqnarray*}
 for $i=1, \ldots, g_1=27$ and $j=1, \ldots, g_2=5$, and where $\delta(i)$ indicates the region to which state $i$ belongs. 
Figure \ref{fig:application1} illustrates samples from this prior under varying parameter conditions, where the different colours indicate regional membership. 
Panels (a) and (d) represent the case where there are differences between regions, but  where the states within each regions are fairly similar. This flexibility would be effectively impossible without an extra level of prior hierarchy.

To handle this extended prior, we modify the functional regression model (\ref{eq:reg.mod}) appropriately by including a regional level predictor. Specifically, the functional regression model for state $i$, located in region $j$, is given by
\begin{align*}
     \label{eq:Ext.reg.mod} \tilde{Z}_{i\ell}(x) = \mbox{offset} + \gamma^i_0(x) + \gamma^i_1(x) \log(\tilde{K}_{i\ell}(x)) + \gamma^i_2(x) \overline{\log}_{R_j}(\tilde{K}_{\ell}(x)) + \gamma^i_3(x) \overline{\overline{\log}}(\tilde{K}_\ell(x)) + \epsilon^\ell(x),
\end{align*}
for $\ell=1,\ldots,m$, where
\begin{align*}
 \overline{\log}_{R_j}(\tilde{K}_{\ell}(x)) = \frac{1}{|R_j|}\sum_{i\in R_j}{\log(\tilde{K}_{i\ell}(x))} \quad \text{and} \quad
 \overline{\overline{\log}}(\tilde{K_\ell}(x)) = \frac{1}{5} \sum_{j=1}^{5}
 \overline{\log}_{R_j}(\tilde{K}_\ell(x)).
\end{align*}
where $R_j$ denotes the set of states in region $j$, and $|R_j|$ the number of states in region $j$.
The functional regressors $\gamma_1^i(x),\gamma_2^i(x)$ and $\gamma_3^i(x)$ correspond to the Gaussian processes in the three levels of the hierarchy: state, region and country. The functional regression-adjustment is then performed by modifying (\ref{eqn:funcregadj}) in the obvious way.  Beyond performing the regression-adjustment, if there is interest in analysing the fitted regression coefficients, then the model may be reparametrised so that  group $i$ is only contained in one of the regressors, as in  (\ref{eq:reg.mod}). Notice that any reparametrisation does not affect the model prediction nor the adjustment.

\begin{figure}[htb]
	\centering
	\subfloat[Kernel density estimates \label{fig:kernels.app}]{\includegraphics[width=6cm,height=6cm,angle=-90]{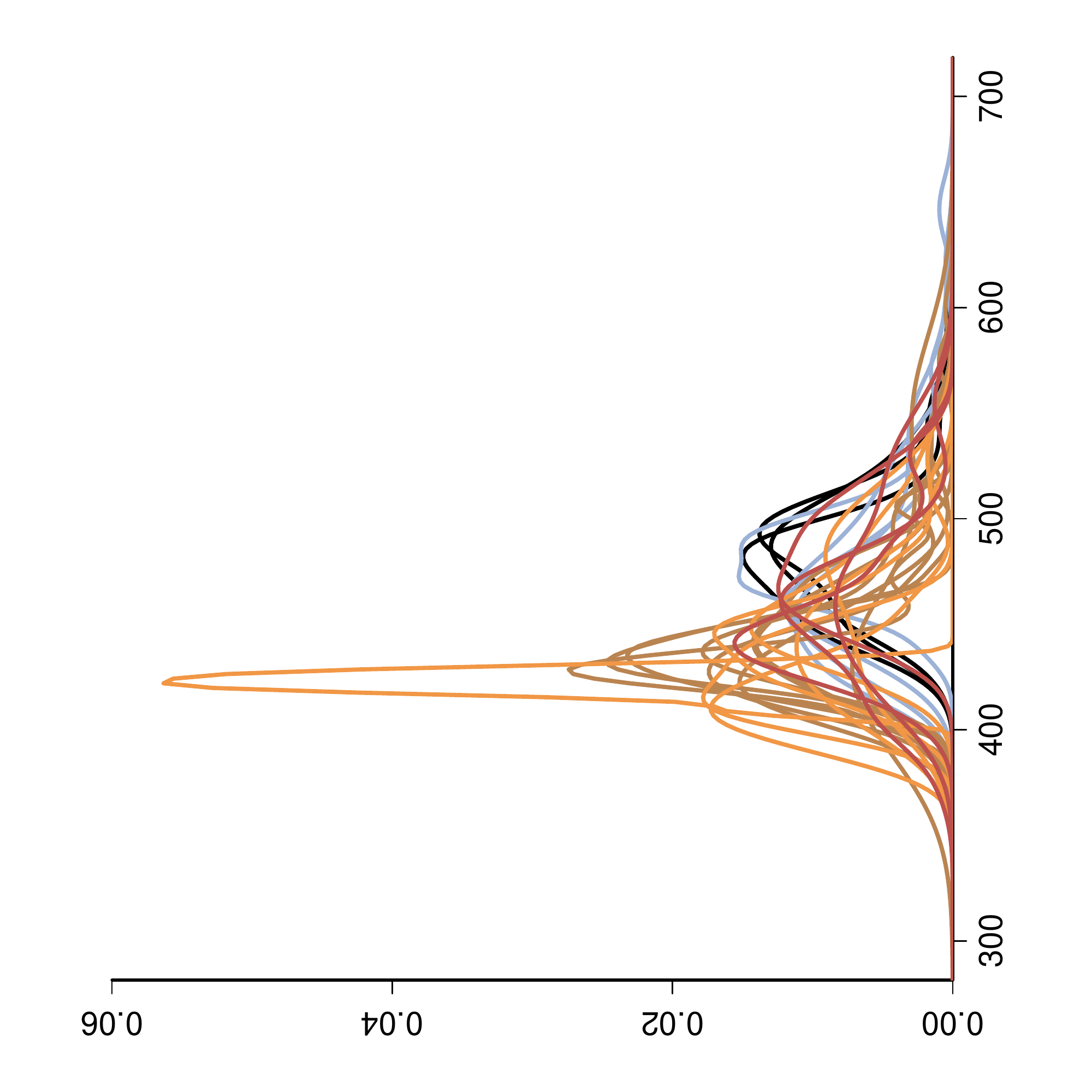}}
	\subfloat[Posterior density estimates \label{fig:post.means.app}]{\includegraphics[width=6cm,height=6cm,angle=-90]{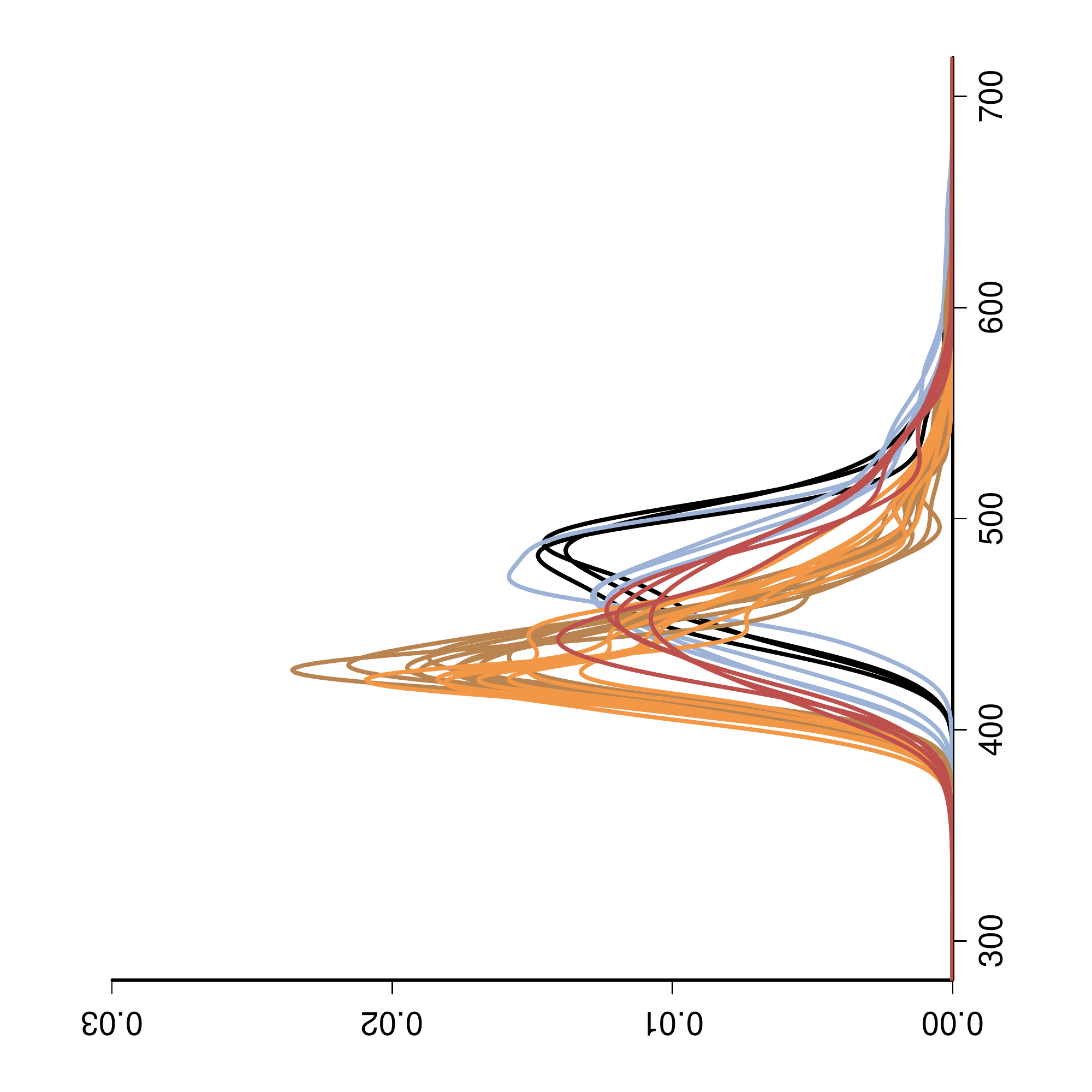}}
	\\
	\subfloat[Sample and ABC means \label{fig:state.means}]{\includegraphics[width=6cm,height=6cm,angle=-90]{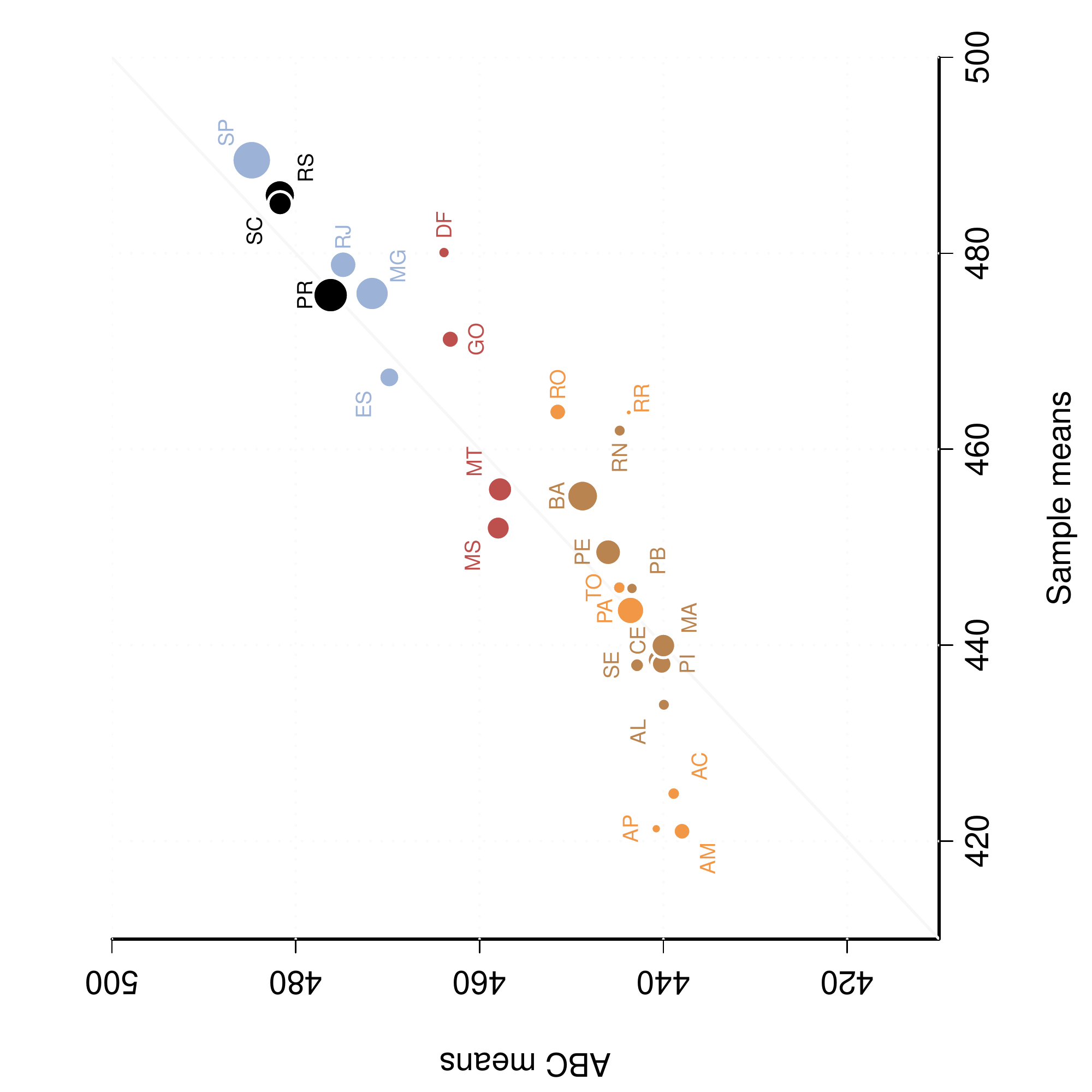}}
	\subfloat[Posterior ranking \label{fig:ranking}]{\includegraphics[width=6cm,height=6cm,angle=-90]{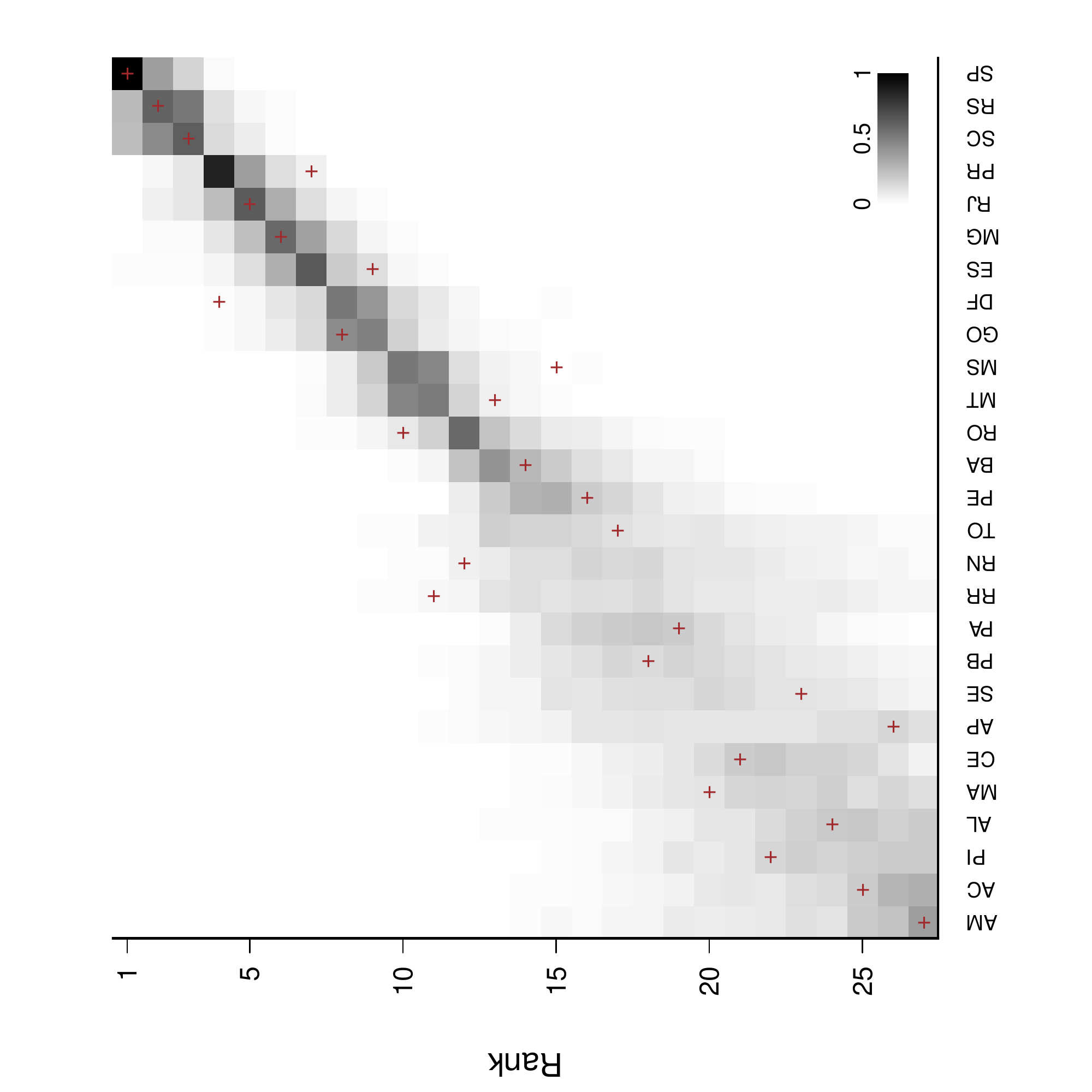}}
	\caption{\small (a) Independent kernel density estimates for each of the $g_1=27$ states in the Brazilian Enem analysis. Line colour indicates region membership. (b) ABC posterior mean density estimates for each state. (c) Sample means versus posterior means for each state. Circle area is proportional to the number of observations in each state. (d) Posterior rank of each state according to the mean of the posterior density estimates. Crosses indicate the rank of each state according to the sample mean ($x$-axis in panel (c)).}
	\label{app.objects}
\end{figure}

In the following, we take the base density $b(x|\phi)$ to be $N(500,5000)$ and specify the prior 
 for the parameters of the covariance functions of the Gaussian processes to be $\sigma_h \sim \mbox{Gamma}(3, 5)$ and $\alpha_h \sim \mbox{Gamma}(1, 10000)$, for $h=1, 2, 3$.
We use $k=200$ grid points and $150$ B-spline basis functions. We generate $10,000$ samples from the prior distribution, and accept the $m=1,000$ samples closest to the observed data.

Figure \ref{app.objects}(a) illustrates independent kernel density estimates of the school's mean grades for each of the 27 states, with line colour indicating membership in the 5 regions. While there is some inter-state variability, there appear to be some visible similarities of density estimates for states in the same region. These similarities become clearer in Figure \ref{app.objects}(b), which shows the regression-adjusted posterior mean density estimates (notice the different plot scale). These densities indicate a strong regional clustering, and some obvious sharing of information between states within each region.

Similarly to the example in Section \ref{sec:simulation}, an analysis of the fitted regression coefficients (not shown) indicates that the functional regressors have contrasting effects on each state. For example, for the heavily populated state of Sao Paulo (SP), the posterior samples are mostly determined by the kernel density estimate for this state. Alternatively, for small states in the Northeast region, such as Alagoas (AL) and Paraiba (PB), the ABC posterior samples are predominantly affected by the average of the kernel estimates of the states within this region. Finally, for the Federal District (DF) and Acre (AC), although not the  dominant regression term, the overall country  kernel estimate plays a substantial role. It is clear that the model is capable of borrowing strength with the appropriate magnitude from the appropriate source for each group.

Figure \ref{app.objects}(c) plots the sample mean for each state against the mean of the mean posterior density estimate, $\mathbb{E}[f_i(x)|\mathcal{D}]$. Here the area of each point is proportional to the sample size within each state. An obvious shrinkage effect can be observed, particularly in states with fewer schools. In addition, the evident regional clusters (indicated by colouring) suggest that geographical forces critically affect the system. 

While not making full use of the posterior density estimates, a simple way to construct a state ranking system could be via the means of these posterior densities.
Figure \ref{app.objects}(d) illustrates the posterior distribution of the rank of each state, according to the mean of the posterior density estimate, with a darker shade indicating greater posterior mass.
 The crosses mark each state's rank according to the independent sample means ($x$-coordinate values in Figure \ref{app.objects}(c)). From the plot, we see that Sao Paulo (SP) is most likely in the top 3 states (probability $>0.98$), and that the Federal District (DF) is likely ranked too high according to the independent sample means, and Mato Grosso do Sul (MS) too low. The posterior rank uncertainty is large for states in the lower part of the ranking. For example, while Roraima (RR) and Amapa (AP) have substantially different sample means (421.27 and 463.76) the posterior probability of the mean of Roraima being greater than the mean of Amapa is considerable at 0.28.  
 
 \begin{figure}[htb]
	\centering
	\subfloat[Sample means \label{fig:map.original}]{\includegraphics[width=8cm,height=8cm,angle=-90]{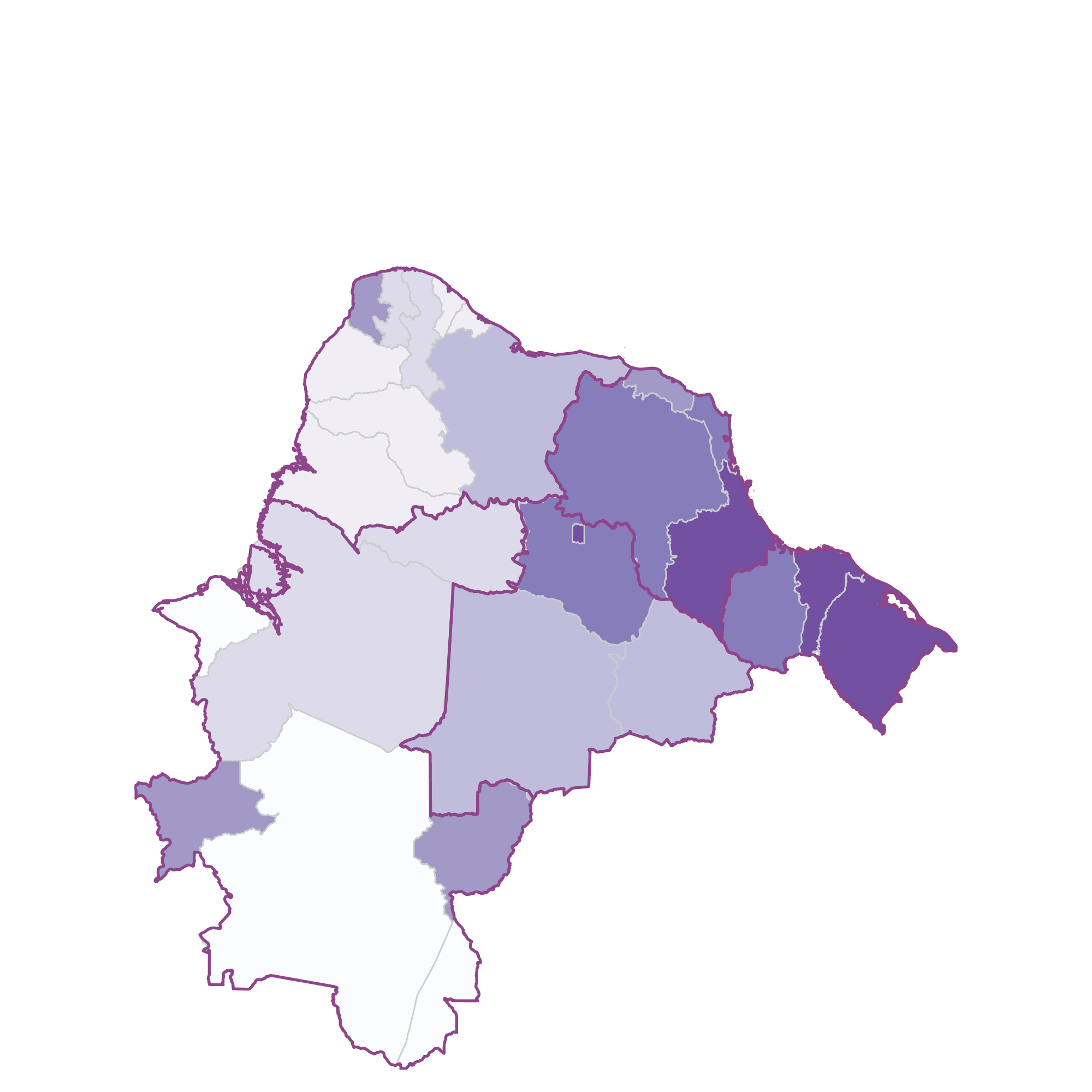}}
	\subfloat[ABC posterior means \label{fig:map.estimated}]{\includegraphics[width=8cm,height=8cm,angle=-90]{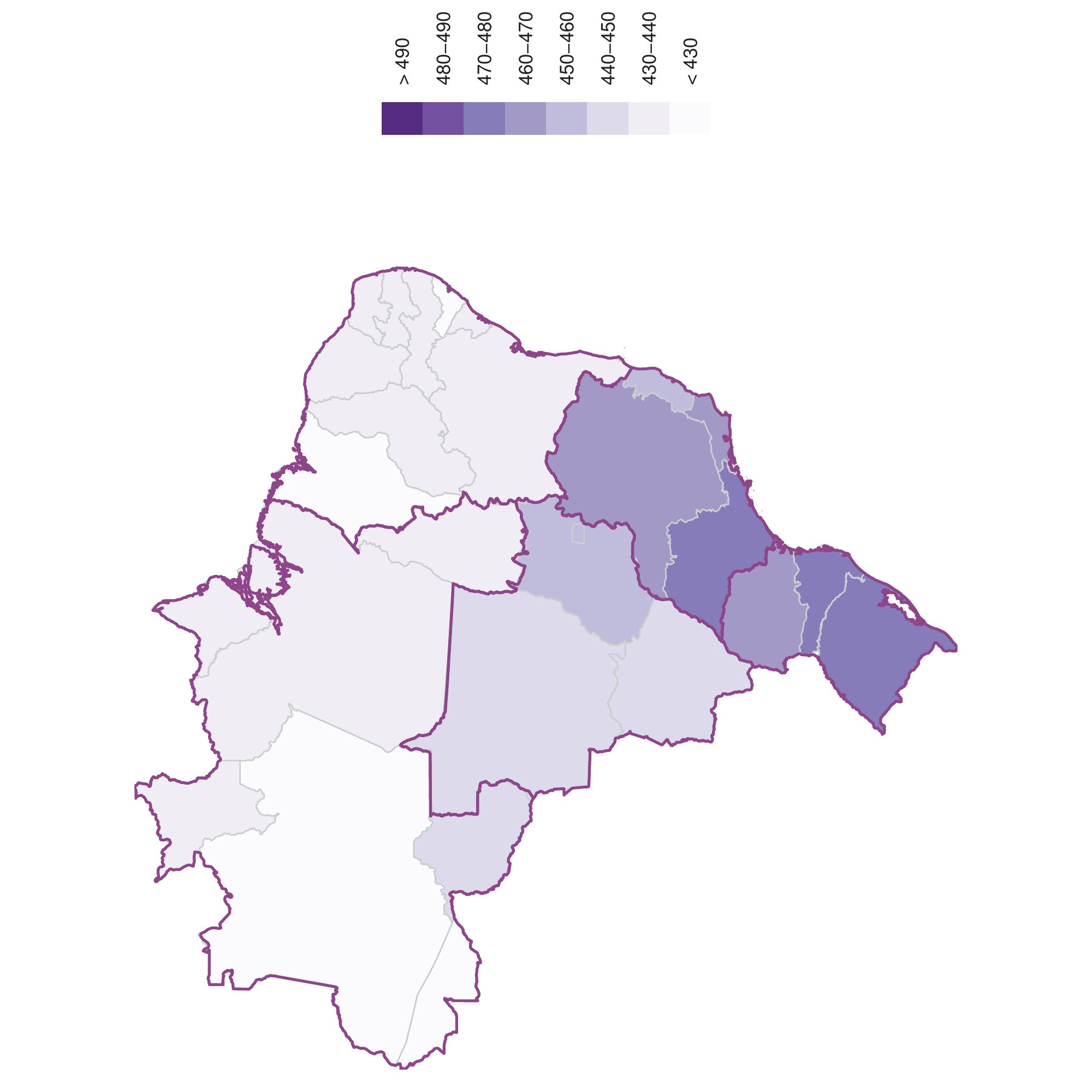}}
	\caption{\small Geographical maps for comparing (a) the sample and (b) posterior density means.}
	\label{fig:maps}
\end{figure}

Finally, Figure \ref{fig:maps} illustrates the spatial distribution of the sample means (panel (a)) and the posterior means (panel (b)). The effect of the hierarchical model is apparent in panel (b), as there is clear evidence of smoothing both within and between regions. There is an obvious north-south effect in the mean performance of schools in the Enem examination. Note that while comparison are made here in terms of means, any other feature could equally be considered, including the densities variability and shape.

\section{Discussion}

The hierarchical Gaussian process prior introduced in this paper permits, in a very simple way, the characterisation of prior beliefs about a set of univariate densities. The use of ABC methods allows the sampling of approximate draws from the posterior distribution at a moderate computational cost. This cost is almost certainly smaller than could be realistically obtained by developing a direct MCMC sampler along the lines of, say,   \cite{Adams2009}.

One advantage of the approach presented here over hierarchical Dirichlet process models is that the parameters in our hierarchical prior are easily interpretable, and allow very direct control over the features of the prior. With Dirichlet process based methods, which build elaborate hierarchies involving mixing distributions, it can be challenging to be expressive of prior information.

The construction of a nonparametric extension to the standard ABC regression-adjustment is a novel contribution to the ABC literature. However, ABC is an approximate inferential procedure which heavily relies on the information content of its summary statistics and in the accuracy of the regression-adjustment model.
Fortunately, there is a range of credible candidates for summarising the data -- kernel density estimation methods have been extensively developed and investigated by numerous authors. Further, regression-adjustment models are well understood in the ABC literature as a method to improve the conditional density estimation of the posterior distribution \citep{blum10}.

Both the simulation study and the analysis of high school exam performance in Brazil highlights the capability of the  ABC procedure to borrow strength across multiple groups, to appropriately produce shrinkage where needed (with small sample sizes), and to suitably handle the boundary-bias problem of the kernel density estimator summary statistic.
Ultimately, the hierarchical Gaussian process density model presents an enhanced approach over existing methodologies, and provides a valuable tool in the practitioners toolbox.

\subsection*{Acknowledgements}

 GSR is funded by the CAPES Foundation via the Science Without Borders program (BEX 0974/13-7).
 DJN is supported by a Singapore Ministry of Education Academic Research Fund Tier 2 grant (R-155-000-143-112).
SAS is supported by the Australia Research Council through the Discovery Project Scheme (DP1092805).

\bibliography{bibli2}

\end{document}